\documentclass[runningheads]{llncs}

\UseRawInputEncoding

\usepackage{latexsym}
\usepackage{paralist}
\usepackage{amssymb}
\usepackage{soul}
\usepackage{enumitem}
\usepackage{mathtools}
\usepackage{xcolor}
\usepackage[all]{xypic}

\usepackage{paralist}

\usepackage{graphics}

\usepackage[belowskip=-15pt,aboveskip=0pt]{caption}

\newtheorem{propos}[example]{Proposition}

\newtheorem{defin}[example]{Definition}

\usepackage{amssymb}

\begin{document}

\title{On human-centred security:  A new systems model based on modes and mode transitions}

\titlerunning{Human-centred security systems}
%

\author{Edwin J Beggs\inst{1}  \and John V Tucker\inst{1} \and Victoria Wang\inst{2}}
\institute{School of Mathematics and Computer Science,\\ Computational Foundry, Swansea University, \\Bay Campus, Fabian Way, \\Swansea, SA1 8EN, United Kingdom\\ \email{e.j.beggs@swansea.ac.uk, j.v.tucker@swansea.ac.uk}\medskip \and School of Criminology and Criminal Justice,\\ Portsmouth University, \\St GeorgeÕs Building, 141 High Street, \\Portsmouth, PO1 2HY, United Kingdom\\ \email{victoria.wang@portsmouth.ac.uk}}
\authorrunning{E.J.\ Beggs, J.V.\ Tucker \& V.\ Wang}
%

\maketitle

\begin{abstract}
We propose an abstract conceptual framework for analysing complex security systems using a new notion of \textit{modes} and \textit{mode transitions}.  A mode is an independent component of a system with its own objectives, monitoring data, algorithms, and scope and limits. The behaviour of a mode, including its transitions to other modes, is determined by interpretations of the mode's monitoring data in the light of its objectives and capabilities -- these interpretations we call \textit{beliefs}.  We formalise the conceptual framework mathematically and, by quantifying and visualising beliefs in higher-dimensional geometric spaces, we argue our models may help both design, analyse and explain systems. The mathematical models are based on  \textit{simplicial complexes}.
\end{abstract}


\section{Introduction} \label{intro}

Security, like safety, is defined by a human context. When studying a software system with a specification, its security is more than a matter of the evaluation of its correctness relative to its specification. What it does in human situations, seen and unforeseen, and the possible consequences matter. Thus, ultimately, security is judged outside the technical world of the system and its specification. Developing this point about security further, of importance for security are critical incidents in which harm to people are possible, immanent or current. The security of a system is intimately connected with expectations and assumptions about the wider human-centred systems it serves: cyber security involves more than cyber.

Here, we address the question:  

\textit{How can we accommodate some human aspects of security systems? Can we develop conceptual frameworks and systematic methods for modelling security systems  and their operation in a human context?}

In seeking some general theory for security, we turn to general system theories and emphasise the role of scenarios in which systems and people relate. General system theory is a loosely defined (huge) portfolio of conceptual frameworks and models  for all sorts of complex systems, whose ideas try to capture: how systems are made up of components; how components cooperate; and how systems and components perform in their respective environments. Common to all complex systems is the role of data about the behaviour of systems and their environments.  Notably,  ideas from general system theory are to be found in our everyday language, as well as in our technical methodologies for designing systems both physical, virtual and human.

Specifically, we offer a new general system model whose central ideas are those of (i) modes of operation, (ii) interpretations of monitoring data and beliefs about the fitness of modes,  (iii) protocols for changing modes, and (iv) the transparency of these features (i)-(iii). Our new model will be described informally by means of working definitions and principles so that the flexibility and scope of its basic concepts can be best appreciated. Then it will formalised by mathematical models that nail down algebraically the relations between the modes as they are determined by data about the environment. In particular, the algebraic models will be quantified and visualised geometrically. 


\subsection{Tools for thinking about security systems}\label{explainability}

Our systems are a mix of equipment, software tools, and people -- the latter organised into teams and also acting as individuals. These are bound together and animated by data and communications. Structurally, complex systems are made of modular and hierarchical components that are autonomous, yet are cooperative, adaptable and responsive to situations -- especially the individuals, depending upon their degree of professional discretion. Any of these components can --  and in practice do  -- fail in some way. The design and performance of such systems in the real world are not easy to specify, measure, predict or audit.   

In security, threats and risks change both in their nature and perceived importance. Significant social and economic disruption can have many causes, such as failures of infrastructures and services, or lack of information. (Extreme weather conditions exemplify such disruptions.) However, our societies are held together by data and software. Thus, cyber failures, and attacks where the objective is to make some software components fail, are always meaningful threats to many aspects of contemporary everyday life, and never far away from national security.  

In this paper, we will address system design in human contexts, almost from first principles. We propose an abstract conceptual framework for analysing systems using a new notion of \textit{mode} and \textit{mode transitions}.  A mode is an independent active component of the system with its own objectives, monitoring data, algorithms, and scope and limits for action. The behaviour of a mode, including its transitions to other modes, is determined by interpretations of the mode's monitoring data and capabilities in the light of its objectives; these we will term \textit{beliefs}.    Crucially, a mode may no longer be fit for purpose and need to be changed to another mode.  A system can be in several modes at once.

We formalise the conceptual framework mathematically and, by visualising the beliefs arising from evaluating a mode's monitoring data in higher-dimensional geometric spaces, we argue our theoretical approach and models may help to explore, design, predict, and explain system behaviour. The various mathematical models are based on mathematical objects called abstract and concrete \textit{simplicial complexes}. 

To demonstrate the framework, we apply our models to three security situations: 

(i) triage for a large set of data about individuals in order to classify `persons of interest';

(ii) mapping the potential causes and effects of a cyber security incident;

(iii) examining a multiagency response to a critical incident, using the UK Gold-Silver-Bronze command structure.

These types of scenario are commonly associated with Tier 1 -- high priority -- risks in national security audits in the UK \cite{NSRA}.


\subsection{Structure of this paper}\label{structure}

This paper has general theoretical aims whose new contributions, and their location, are:

1. To reflect on scenarios and their part in the security systems specification (Section \ref{security_scenarios} and \ref{Potential}). 
 
2. To define general concepts and principles for a conceptual framework to model systems using ideas about modes and mode transitions (Section \ref{modes}).

3. To create geometric structures to visualise changes of beliefs about the behaviour of systems over time (Section \ref{modes}). 

4. To turn the conceptual framework into a rigorous mathematical theory that enables quantification and evaluation of the reasons and beliefs behind decision making (Section \ref{modes} and the Appendix).

5. To apply these general methods in three security scenarios (Section \ref{illus}).

6. To reflect on possible next steps (Section \ref{Potential}).

In the matter of 5, we explain the mathematical ideas through examples; the general mathematical definitions of simplicial complex and belief function are given in the Appendix. A general mathematical introduction to modes and their theory is our \cite{BeTuSimp}; there, our early thinking about modes was shaped by modelling autonomous \textit{physical} systems.


\subsection{Acknowledgments}
The authors would like to acknowledge the support by UK's  Accelerated Capability Environment ACE-C391 National Security Tech Surprise.

\section{Human-centred security systems}\label{security_scenarios}

To begin, we observe that human aspects of security are commonly expressed through scenarios.

\subsection{Responding to security incidents}

Central to security are scenarios for critical incidents against which the security of the system can be explored. By a \textit{security scenario} we have in mind (i) a hypothetical situation that involves potential threats and vulnerabilities or (ii) a description of a real-life situation, past or present, that manifested threats and revealed vulnerabilities.  Of particular importance are scenarios containing critical incidents.
A \textit{critical incident} creates an outcome or consequence that is of significant harm to an individual, community, or business, or to public confidence, broadly interpreted. 
Any critical incident arising from a system's operation  calls for one or more other systems to respond and contain damage, make repairs and restore operation.

A security scenario should contain information necessary to identify parameters that enable a relevant actor to simulate potential threats effectively, with the aim of being ready to prevent, prepare for, detect and respond to critical security incidents. These scenarios typically belong to security professionals, staff and managers of organisations, and policy makers. Scenario-based research is common in cybersecurity. For example, the major role scenarios play in cybersecurity testbed development, as well as in cybersecurity testing for experimental and educational purposes has been thoroughly discussed in the systematic literature review \cite{Yamin_et_al2020}.

Scenarios are themselves classified using ideas about risk, e.g., likelihood, impact and mitigation. For example, the \textit{UK National Security Strategy} \cite{CabinetOffice2010} is based on the assessment of many security scenarios. In the methodology section, the document comments the ``plausible worst-case scenario of each risk was then scored in terms of its likelihood and its potential impact. In order to compare the likelihood of one risk against another and to make relative judgements, these plausible worst-case scenarios were plotted on a matrix  (\cite{CabinetOffice2010}, p. 4).


\subsection{Critical incidents and the Gold-Silver-Bronze command structure of the UK }\label{GSB}

In responding to a critical incident, many types of agencies and organisations, with their specialist units and systems, may be called upon to collaborate and cope with an unknown and unfolding situation. Concepts and working principles, based on experience, are needed for this to work effectively.  Such frameworks must `cage the incident' in order to initialise and develop a response; they must also be known and understood by the participating agencies and organisations. In the UK, such frameworks for a diverse set of government agencies and services have been created, used and exercised for decades. To pick one example, relevant to a later case study, a glimpse of these frameworks can be found in the Home Office's \textit{Critical incident management for staff of Border Force, Immigration Enforcement and UK Visas and Immigration} \cite{HomeOffice2021}.

In the UK, the \textit{Gold-Silver-Bronze (GSB) command structure} is used for major operations by the emergency services \cite{HomeOffice2021}. The structure was originally created by the Metropolitan Police Service, as a direct response to the Broadwater Farm Riot on the 6 October 1985 \cite{BBCNews2015}. Assessing their responses during this riot, the Metropolitan Police came to the realisation that their traditional rank-based command system was inappropriate and inefficient in dealing with sudden events. (Particularly, on the night of the riot, it was not clear who was actually in charge of the actual operational procedures of the police.) 

Thus, the GBS is a role-based system that assigns three different level of command according to skill, expertise, location and competency.\footnote{In some cases, the national government (via the Cabinet Office Briefing Rooms) assume ultimate control and act as a fourth, platinum level (Home Office, 2021).} The three levels are \textit{strategic}, \textit{tactical} and \textit{operational command structures}.

Briefly, the Gold Commander is responsible for the strategic direction and has overall control of the resources at the incident. Instead of on site, the Gold Commander is in a distant control room formulating the strategy for dealing with the incident. 

Silver Commanders are the seniors in charge of their own organisations resources available at the scene. They are responsible for the tactical coordination, deciding on how to use their resources to achieve the strategic aims of the Gold Commander. Like the Gold Commander, they are also not directly involved in dealing with an incident itself. At the scene of the incident, they will work closely with the Silver Commanders of other organisations, operating from a purpose-built command vehicle or makeshift command room(s), known as the Joint Emergency Services Control Centre (JESCC). 

The Bronze Commanders are responsible for the operational implementation, directly controlling their organisations resources at the incident and working with colleagues at the scene of the incident. Bronze Commanders may share responsibilities and tasks in complex incidents and assume responsibility for different areas, if an incident is widespread geographically. 

During the initial stages of an incident, the first member of the first organisation to arrive at the incident temporarily assumes the role of Silver or Bronze Commander until relieved by a more senior member of their organisation. It is noteworthy that these three roles are not restricted to ranks, though invariably the chain of command will follow the order of rank.

Although the GSB command structure was originally devised to respond to sudden major incidents, it spread-out in all police forces and emergency services, and has been frequently used in pre-planned operations (e.g., the policing of football matches or firearms operations (cf. \cite{HomeOffice2021}).


\section{Modes and their visualisation} \label{modes}

Modes decompose and categorise the various objectives and behaviours of a system.   
A mode collects  relevant data and applies algorithms to govern the behaviour and external communication. 


\subsection{What is a mode?}\label{what_is_ a_mode}

Here is a working definition to begin building the conceptual framework:
 
\begin{definition}\label{mode_working definition}
Consider a system $S$ operating in an environment $E$.
A {\em mode} of the system $S$ is defined by these characteristics:

\noindent
1.  A mode is associated to, or determines, a subset of the possible states of the system.

\noindent
2.  A mode is designed to deliver on objectives for the system when in these states.

\noindent
3. A mode consists of

(a) methods to input data from the environment $E$ and to output data to $E$.

(b) data types and algorithms for implementing the objectives.

\noindent
4. A mode has means to evaluate its performance against its objectives and, if necessary, choose and transfer to another mode.
\end{definition}

Thus, a mode is responsible for specific aspects of the system's performance. These may be high-level, primary objectives and purposes of the system, or lower-level technical services. It may be autonomous or a platform for accessing other modes. Each mode owns its relationship with the environment.


\subsection{Principles for designing modes}\label{basic_principles}

With the initial intuitions of Definition \ref{mode_working definition} in mind,  we can formulate some design principles to develop a conceptual framework for thinking about systems in terms of modes and mode transitions.

\medskip\noindent
\textbf{Completeness}. A set of modes for a system is a classification of the operation or behaviour of a system. At any time, a system can be in one, or more, modes. 

\medskip\noindent
\textbf{Composition}. When a system is in a number of modes then that situation itself constitutes a mode.

\medskip\noindent
\textbf{Component}. A set of modes for a system consists of (i) a set of \textit{basic modes} and (ii) \textit{joint modes} made by combining other modes. 

\medskip\noindent
\textbf{Localisation}. Each mode possesses its own data to monitor its behaviour and environment. This monitoring data determines a local state space called the \textit{evidence space} of the mode that represents what its modes can know.

\medskip\noindent Thus, knowledge of the system at any time is localised to the modes at that time.

\medskip\noindent 
\textbf{Globalisation}. What the system can know about its environment is a synthesis of what its modes can know of the environment through their monitoring data.  By combining the evidence spaces, an idealised global state space for the system is possible for reasoning.

\medskip\noindent
\textbf{Quantification}. If a state of the system is meaningful for a number of modes then the relevance or suitability of these modes for the state of the system must be quantified, calibrated and interpreted.

\medskip\noindent
\textbf{Visualisation}.
With quantification and calibration comes the possibility of visualisation via geometric objects drawn to scale and via derived qualitative diagrams.

\medskip\noindent
Shortly, we will show how to visualise in space the basic modes by vertices, and the joint modes arising by combining them as lines, triangles, tetrahedra etc. The geometric objects we build are called \textit{simplicial complexes}. Simplicial complexes are made up of geometric pieces called \textit{faces}. 

\medskip\noindent
\textbf{Modes and faces}. In a successful system model using modes, \textit{every face of its simplicial complex is a mode and every mode is represented by a face}.

\medskip\noindent
Quantification can take the form of a position in the space representing the suitability of the mode for the state of the system, in a line or triangle or tetrahedron, etc.\ representing the joint mode.

\medskip\noindent
\textbf{Belief}.  The position on a face is computed by \textit{belief functions} that from a state of the system calculates a measure of the relevance of other modes to that state.

\medskip\noindent
In the time evolution of the system we need to decide

(i) if, and when, a system should change from one mode to another, and 

(ii) which new mode should chosen.

\medskip\noindent
\textbf{Thresholds}. The transition out of one mode into another is governed by the results of the quantification and calibration. The decision to move to a new mode may be specified by numerical thresholds. Transition has these stages:

(a) the realisation that the mode is approaching its limitations

(b) the selection of modes that could be more appropriate

(c) triggers to choose and change to a new mode.

\noindent The belief functions, in computing relevance, are a means to trigger changes of mode.

\medskip\noindent
\textbf{Explanation}. The conceptual system of modes, belief functions, mode evaluation, thresholds and mode transition functions, and their visual representation in simplicial complexes, can serve as an explanatory framework for the dynamical behaviour of automatic systems.

In particular, the position and path over time (trajectory) in the simplicial complex visualises the dynamics of decision making by the system.


\subsection{Simple security examples}\label{examples}
 
 We consider some simple examples to begin to shape our informal ideas of a system of modes, the evaluation of beliefs and how they might be visualised. 
 
  \begin{figure}[htbp]
\begin{center}
 \unitlength 0.45 mm
\begin{picture}(130,51)(10,27)
\linethickness{0.3mm}
\put(20,35){\line(0,1){30}}
\linethickness{0.3mm}
\multiput(70,50)(0.24,0.12){167}{\line(1,0){0.24}}
\linethickness{0.3mm}
\put(110,30){\line(0,1){40}}
\linethickness{0.3mm}
\multiput(70,50)(0.24,-0.12){167}{\line(1,0){0.24}}
\linethickness{0.2mm}
\put(70,50){\line(1,0){37}}
\linethickness{0.2mm}
\put(113,50){\line(1,0){12}}
\linethickness{0.3mm}
\multiput(110,70)(0.12,-0.16){125}{\line(0,-1){0.16}}
\linethickness{0.3mm}
\multiput(110,30)(0.12,0.16){125}{\line(0,1){0.16}}
\linethickness{0.1mm}
\multiput(82,48)(0.22,0.12){117}{\line(1,0){0.22}}
\linethickness{0.1mm}
\multiput(92,44)(0.18,0.12){83}{\line(1,0){0.18}}
\linethickness{0.1mm}
\multiput(112,57)(0.12,-0.13){67}{\line(0,-1){0.13}}
\linethickness{0.1mm}
\multiput(112,48)(0.12,-0.12){42}{\line(1,0){0.12}}
\linethickness{0.1mm}
\multiput(100,40)(0.14,0.12){50}{\line(1,0){0.14}}
\linethickness{0.3mm}
\multiput(89,57)(0.24,0.12){42}{\line(1,0){0.24}}
\put(99,62){\vector(2,1){0.12}}
\linethickness{0.3mm}
\multiput(89,43)(0.24,-0.12){42}{\line(1,0){0.24}}
\put(99,38){\vector(2,-1){0.12}}
\linethickness{0.3mm}
\put(90,52){\line(1,0){10}}
\put(100,52){\vector(1,0){0.12}}
\linethickness{0.3mm}
\put(23,44){\line(0,1){10}}
\put(23,44){\vector(0,-1){0.12}}
\put(20,70){\makebox(0,0)[cc]{judgement of monitoring data}}

\put(20,30){\makebox(0,0)[cc]{intervention}}

\put(55,54){\makebox(0,0)[cc]{incoming}}

\put(56,46){\makebox(0,0)[cc]{calls}}

\put(128,70){\makebox(0,0)[cc]{police}}

\put(130,30){\makebox(0,0)[cc]{ambulance}}

\put(135,51){\makebox(0,0)[cc]{fire}}

\put(-16,54){\makebox(0,0)[cc]{triggering}}
\put(-16,46){\makebox(0,0)[cc]{an action}}

\put(174,54){\makebox(0,0)[cc]{triage for}}
\put(174,46){\makebox(0,0)[cc]{emergency services}}

\end{picture}
\setlength{\belowcaptionskip}{-25pt}
\medskip
\caption{Visualising a mode transition for a trigger mechanism and triage for emergency services}
\label{modpic}
\end{center}
\end{figure}
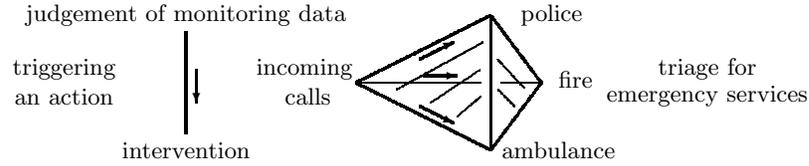

\begin{example}[\textbf{A simple trigger}]\label{man58}
 In Figure~\ref{modpic}, the left diagram represents visually a trigger for action that has three modes: two vertices and one line. Consider the line between a mode that makes judgements of some attribute using data from monitoring the environment (the point at the top of the interval) and the mode that performs an intervention (the point at the bottom). Unlike a usual state transition graph, where lines merely represent transitions or jumps, here the line \textit{itself} represents a mode where the decision to intervene unfolds according to its monitoring data. These observations are formed into a numerical quantification of beliefs about the state of the system and visualised as points along the line, beginning at the top and moving to the bottom. Threshold points on the line begin to warn, and then trigger, a change of mode; on reaching the bottom, the intervention begins. 
 
A witness can note the position along the line to see how things were going, and perhaps be reassured that the course of action was progressing correctly.
Such a point on a line indicator is quite a common visualisation (e.g., as with a progress bar that simply estimates time taken for a task to be completed). Here, however, the line indicator is displaying the an interpretation of information which the control system is actually using to make decisions. If the point is near the top, then a witness knows that the control system has no intention of intervening because they are looking directly at the system's own assessment of the situation. 
\qed
\end{example}
 
In this example, we see that a key to the transparency of decision making is continuity. We do not discontinuously move between vertices on a graph, but have a continuous motion in a geometric visualisation, which allows  \textit{an estimate of what the system is about to do and when}. We make a general comparison between visualising system change using complexes versus graphs in the Appendix, Section \ref{gra7}.

Of course, another key to transparency is reliability, that what we see is really related to the state of the system.  This means that \textit{the geometry of the visualisation must be  intimately related to how the system actually makes decisions}.

\begin{example}[\textbf{A simple emergency}]\label{man59}
In the second diagram of Figure~\ref{modpic}, we consider a notification that warns of a possible incident -- such as personal attack, social disturbance, or collision of vehicles.  In general, the incident could involve some or all of the primary first responders of police, fire brigade and ambulance. These are represented by vertices in the diagram.  As more information becomes available over (hopefully a short) time period, relevant choices of services can be made and deployed to the scene. In particular, between the separate actions of say deploying police and ambulance we have a combined action of doing both simultaneously, and that is represented in the diagram by a line between the two vertices.

Altogether the 15 faces of the diagram represent 15 modes of the system!
The filled-in tetrahedron between the four vertices is the mode which gathers all available information and assesses the response. When beliefs about the state of the system (represented by a point in this tetrahedron or 3-simplex) reaches the triangle at the end then full deployment is instigated. If the incident in some way is misrepresented then the point moves, possibly returning to the warning vertex, and then to the rest of the system (not shown). The state when an action is actually taken should be one of the three action vertices two of the lines between them. 
\qed
\end{example}

\subsection{Geometric intuitions}\label{general}

Systems frequently have to take account of several factors or carry out several tasks, and we can take account of this by allowing combinations of modes to also form modes, the joint modes of Section \ref{basic_principles}. These joint modes translate into edges, triangles, tetrahedra, etc.
In general, we shall represent beliefs about the state of a system as a point in a face of a simplicial complex. What exactly is a simplicial complex? 

A 0-simplex is just a point, and a 1-simplex is a line between some of the 0-simplices. Thus, a simplicial complex consisting of only 0 and 1-simplicies is just a graph with edges the 1-simplices and vertices the 0-simplices. We shall not assume that this graph is directed -- if there are arrows they are either imposed by principles (e.g.,\ not being able to stop an intervention once triggered) or by the current objectives. 
Increasing the dimension, a 2-simplex is a filled in triangle and a 3-simplex is a solid tetrahedron, and so on. Fully general definitions are given in the Appendix.

Consider the visualisation of the simple emergency with its 15 faces: there are 4 0-simplicies, 6 1-simplicies, 4 2-simplicies, and 1 3-simplex. Each of these is a mode.

The belief about the state of the system is visualised as a point in the geometric simplicial object, the position in a face shows which modes the system is in and which mode is likely to come next. The position is given by the belief generated by algorithms on the basis of evidence. Technically, there is mathematical belief theory that is a generalisation of probability theory and, again, we refer to the Appendix for a more formal discussion. 

As the modes are linked to the behaviour of the system, the visualisation of the modes gives the current behaviour of the system, at least its intentions and its capabilities. 


\section{Illustrative security scenarios and systems}\label{illus}

We now apply these ideas to three examples. We describe the first in some detail, including its mathematics; the others are less detailed due to their more open-ended nature and limitations of space.

\subsection{Triage: Persons of interest} \label{illust}

Consider persons coming to the attention of the police as potential security risks through an initial assessment process or triage.

\medskip
\noindent\textbf{State space and design.} 
In the modal design process we first start with our `state space', which will be the set of people in a country, together with their actions and beliefs.
To be more exact, we take a cover of the set of
`people of interest' in the population. 
We wish to classify those into the following subsets:

\medskip
\noindent $PeopleofInterest$ (PoI) -- those who have come to the attention of the security services

\noindent $BeginTriage$ -- those who are in process of classification by the triage process

\noindent $Opportunity$ -- those who are in a position to cause damage by, e.g., access to public platforms or proximity to high value targets.

\noindent $Concern$  -- those who have extreme views or contacts which could pose a security concern.

\noindent $SecurityClearance$  -- those who have been granted clearance for certain activities

\noindent $FurtherInvestigation$  -- those who are under active investigation by the security services

\medskip
These and other subsets of the population are assigned as basic modes. It is clear that a person may be in several of these subsets, whence they belong to joint modes.  For example, an individual who is in both $Opportunity$ and $Concern$ is a larger threat than a person who is in just one of the subsets. 

Now, we consider how to determine whether a person is actually in one of these subsets - this is done by the \textit{oracle calls} mentioned earlier. This could include examining criminal or financial records or interrogation. This process results in a \textit{space of evidence}, and using this we calculate
 the belief functions which we use choose which subsets we believe a person belongs to. These beliefs are continually updated as, e.g.,\ an investigation comes to an end or new evidence is presented.
The belief functions can then then trigger a change of mode, e.g.,\ moving to a security clearance process. 
Of course, the computed belief functions are only an approximation to the true state of affairs, as any process is error prone.

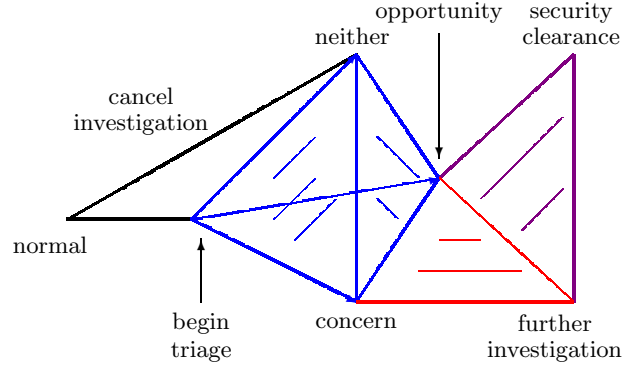
\begin{figure}[htbp]
\begin{center}
\unitlength 1.1 mm
\begin{picture}(76,47)(12,10)
\linethickness{0.3mm}%
{\color{black}{\put(15,30){\line(1,0){15}}%
\multiput(15,30)(0.21,0.12){167}{\line(1,0){0.21}}%
}}%
{\color{blue}{\linethickness{0.3mm}%
\multiput(30,30)(0.12,0.12){167}{\line(1,0){0.12}}
\put(50,50){\vector(1,1){0.12}}
\linethickness{0.3mm}%
\multiput(50,50)(0.12,-0.18){83}{\line(0,-1){0.18}}
\linethickness{0.3mm}
\multiput(50,20)(0.12,0.18){83}{\line(0,1){0.18}}
\linethickness{0.3mm}
\put(50,20){\line(0,1){30}}
\linethickness{0.3mm}
\multiput(30,30)(0.24,-0.12){83}{\line(1,0){0.24}}
\put(50,20){\vector(2,-1){0.12}}
\linethickness{0.3mm}%
\linethickness{0.15mm}
\multiput(30,30)(0.71,0.12){42}{\line(1,0){0.71}}
\put(60,35){\vector(4,1){0.12}}
}}%
{\color{violet}{\linethickness{0.3mm}%
\multiput(60,35)(0.13,0.12){125}{\line(1,0){0.13}}%
\linethickness{0.3mm}
\put(76.25,20){\line(0,1){30}}%
}}%
{\color{red}{\linethickness{0.3mm}%
\put(50,20){\line(1,0){26.25}}
\linethickness{0.15mm}
\multiput(60,35)(0.13,-0.12){125}{\line(1,0){0.13}}%
}}%
{\color{blue}{%
\linethickness{0.05mm}%
\multiput(40,35)(0.12,0.12){42}{\line(1,0){0.12}}%
\linethickness{0.05mm}
\multiput(40,30)(0.12,0.12){42}{\line(1,0){0.12}}
\linethickness{0.05mm}
\multiput(42.5,27.5)(0.12,0.12){42}{\line(1,0){0.12}}
\linethickness{0.05mm}
\multiput(52.5,40)(0.12,-0.12){42}{\line(1,0){0.12}}
\linethickness{0.05mm}
\multiput(52.5,32.5)(0.12,-0.12){21}{\line(1,0){0.12}}%
}}%
{\color{violet}{%
\linethickness{0.05mm}%
\multiput(65,32.5)(0.12,0.12){83}{\line(1,0){0.12}}
\linethickness{0.05mm}
\multiput(70,28.75)(0.12,0.12){42}{\line(1,0){0.12}}%
}}%
{\color{red}{%
\linethickness{0.05mm}%
\put(60,27.5){\line(1,0){5}}
\linethickness{0.05mm}
\put(57.5,23.75){\line(1,0){12.5}}%
}}%
\linethickness{0.15mm}%
\put(60,37.5){\line(0,1){15}}
\put(60,37.5){\vector(0,-1){0.12}}
\linethickness{0.15mm}
\put(31.25,20){\line(0,1){7.5}}
\put(31.25,27.5){\vector(0,1){0.12}}

\put(76,55){\makebox(0,0)[cc]{security}}

\put(76,52.25){\makebox(0,0)[cc]{clearance}}

\put(74,17.5){\makebox(0,0)[cc]{further}}

\put(74,14){\makebox(0,0)[cc]{investigation}}

\put(50,17.5){\makebox(0,0)[cc]{concern}}

\put(60,55){\makebox(0,0)[cc]{opportunity}}

\put(31.25,17.5){\makebox(0,0)[cc]{begin}}

\put(31.25,14){\makebox(0,0)[cc]{triage}}

\put(13,27){\makebox(0,0)[cc]{normal}}

\put(49.5,52.25){\makebox(0,0)[cc]{neither}}

\put(24,45){\makebox(0,0)[cc]{cancel}}
\put(24,41.5){\makebox(0,0)[cc]{investigation}}

\end{picture}
\setlength{\belowcaptionskip}{-25pt}
\caption{Initial assessment for potential security risks}
\label{sectri}
\end{center}
\end{figure}

\medskip
\noindent\textbf{Scenario and its visualisation.} Our scenario is sufficiently simple to be described directly using modes and their geometric representation: consider Fig.~\ref{sectri}. We begin at the `begin triage' vertex of the blue tetrahedral mode `triage'. After conducting an initial investigation, `triage' outputs either (i) a value along the edge from `opportunity' (i.e.,\ there is an opportunity to carry out an attack, e.g.,\ the subject is in close proximity to potential targets) to `concern' (i.e.,\ there is evidence that the subject may be a security risk), or (ii) the vertex `neither' of the above.

\medskip
From `neither' the investigation is cancelled, deleting as much material as procedures allow. From the vertex `opportunity' on the edge (i.e.,\ there is minimal concern about a security risk) the purple triangle `obtain security clearance' mode is started. This will get references from contacts of the subject etc.
From anywhere else on the edge (i.e.,\ there is \textit{some} concern about security risk) the red `consider further investigation' mode is entered. It is possible to move between the `obtain security clearance' and `consider further investigation' modes if new evidence is uncovered. The end points are either being granted `security clearance'  or being entered for a full `further investigation'. To repeat, there are four outcomes and we can leave the triage mode at three vertices or \textit{at one edge}. 

Inspecting the diagram, there are 7 0-simplices, 12 1-simplices, 7 2-simplices, and 1 3-simplex. Thus, the model reveals that there are, in principle, 27 modes in this scenario.

\medskip
\noindent\textbf{Quantification: Space of evidence and thresholds.} 
The above informal description is now quantified and modelled mathematically. We will build a relevant belief function to plot points based on the evidence available; these points lie in the blue triage tetrahedron. 

First, we must mathematically model the space of evidence. This will be based upon concern and opportunity.
We begin with choosing positive numbers 
$$(x_{\mathrm{begin}}, x_{\mathrm{con}}, x_{\mathrm{opp}}).$$ 

\noindent These numbers are ratios having the general form of 
 \[
 \frac{ \mathrm{individual\ score} }{\mathrm{preset\ score} }.
 \]
In this case the numbers try to measure concern $x_{\mathrm{con}}$ and opportunity $x_{\mathrm{opp}}$, and they are estimated as follows.

We take $x_{\mathrm{con}}\ge 0$ to be a weighted sum of evidence from phone calls, emails and social media posts etc., divided by a preset `level of concern' and capped at 1; so $x_{\mathrm{con}}= 1$ is taken to be a significant concern and $x_{\mathrm{con}}= 0$ is no concern.

We set $x_{\mathrm{opp}}$ to be a sum of events attended by the person, weighted by the profile of the event or ease of attack, again divided by a preset level and capped at 1; so $x_{\mathrm{opt}}= 1$ is taken to be a significant opportunity. For instance, a large number of applications to certain events would also be treated as a cause for concern, so these numbers are not independent. 

We might assign
$x_{\mathrm{begin}}\in[0,1]$ as the fraction of the checks remaining to be made; so if we had 54 checks in total to perform we would begin at $54/54=1$ and then count down as checking tasks were completed. However, we may cut short the process if we already have enough evidence to escalate the investigation. We now define
\[
x_{\mathrm{end}} = \max\big\{ 1-x_{\mathrm{begin}},x_{\mathrm{con}},x_{\mathrm{opp}}  \big\}.
\]

\noindent Our space of evidence is this cube of triples:
$$(x_{\mathrm{opp}},x_{\mathrm{con}},x_{\mathrm{end}})\in [0,1]^3.$$

 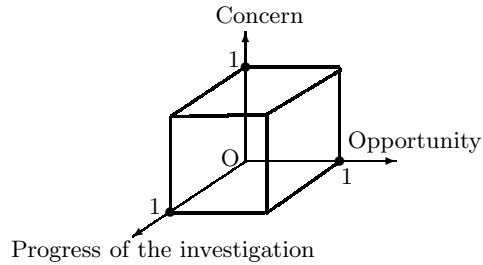
\begin{figure}[htbp]
\begin{center}

\unitlength 0.5 mm
\begin{picture}(80,66.25)(0,10)
\linethickness{0.15mm}
\put(40,40){\line(0,1){34.38}}
\put(40,74.38){\vector(0,1){0.12}}
\linethickness{0.15mm}
\multiput(10,20)(0.18,0.12){167}{\line(1,0){0.18}}
\put(10,20){\vector(-3,-2){0.12}}
\linethickness{0.15mm}
\put(40,40){\line(1,0){40}}
\put(80,40){\vector(1,0){0.12}}
\linethickness{0.3mm}
\put(40,65){\line(1,0){25}}
\linethickness{0.3mm}
\put(65,40){\line(0,1){25}}
\linethickness{0.3mm}
\put(20,26.88){\line(0,1){25}}
\linethickness{0.3mm}
\put(20,26.25){\line(1,0){25}}
\linethickness{0.3mm}
\multiput(45.62,26.25)(0.17,0.12){115}{\line(1,0){0.17}}
\linethickness{0.3mm}
\multiput(20,51.88)(0.18,0.12){109}{\line(1,0){0.18}}
\linethickness{0.3mm}
\multiput(20,51.88)(5.12,0.12){5}{\line(1,0){5.12}}
\linethickness{0.3mm}
\multiput(45.62,52.5)(0.2,0.12){99}{\line(1,0){0.2}}
\linethickness{0.3mm}
\put(45.62,26.25){\line(0,1){26.25}}
\put(43.75,79){\makebox(0,0)[cc]{Concern}}

\put(85,45){\makebox(0,0)[cc]{Opportunity}}

\put(18.12,16){\makebox(0,0)[cc]{Progress of the investigation}}

\put(67,36){\makebox(0,0)[cc]{1}}

\put(37,67){\makebox(0,0)[cc]{1}}

\put(16,28){\makebox(0,0)[cc]{1}}

\put(36,41){\makebox(0,0)[cc]{O}}

\put(55,13.12){\makebox(0,0)[cc]{}}

\put(40,65){\makebox(0,0)[cc]{$\bullet$}}

\put(65,40){\makebox(0,0)[cc]{$\bullet$}}

\put(20,26.25){\makebox(0,0)[cc]{$\bullet$}}

\end{picture}

\caption{The axes for the space of evidence for the investigation}
\label{spaceofevid}
\end{center}
\end{figure}

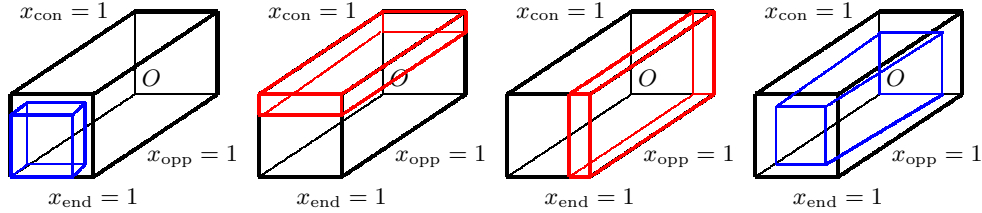
\begin{figure}[htbp]
\begin{center}
\unitlength 0.55 mm
\begin{picture}(60,55)(95,0)

\linethickness{0.4mm}
\put(10,10){\line(0,1){20}}
\linethickness{0.4mm}
\put(10,30){\line(1,0){20}}
\linethickness{0.4mm}
\put(30,10){\line(0,1){20}}
\linethickness{0.4mm}
\put(10,10){\line(1,0){20}}
\linethickness{0.4mm}
\multiput(10,30)(0.18,0.12){167}{\line(1,0){0.18}}
\linethickness{0.4mm}
\multiput(30,30)(0.18,0.12){167}{\line(1,0){0.18}}
\linethickness{0.4mm}
\put(40,50){\line(1,0){20}}
\linethickness{0.4mm}
\put(60,30){\line(0,1){20}}
\linethickness{0.4mm}
\multiput(30,10)(0.18,0.12){167}{\line(1,0){0.18}}
\linethickness{0.1mm}
\multiput(10,10)(0.18,0.12){167}{\line(1,0){0.18}}
\linethickness{0.1mm}
\put(40,30){\line(0,1){20}}
\linethickness{0.1mm}
\put(40,30){\line(1,0){20}}
\put(23,50){\makebox(0,0)[cc]{$x_{\mathrm{con}}=1$}}
\put(54,15){\makebox(0,0)[cc]{$x_{\mathrm{opp}}=1$}}
\put(44,34){\makebox(0,0)[cc]{$O$}}
\put(30,5){\makebox(0,0)[cc]{$x_{\mathrm{end}}=1$}}

{\color{blue}
\linethickness{0.3mm}
\put(10,25){\line(1,0){15}}
\linethickness{0.3mm}
\put(25,10){\line(0,1){15}}
\linethickness{0.3mm}
\put(10,10){\line(1,0){15}}
\linethickness{0.3mm}
\put(10,10){\line(0,1){15}}
\linethickness{0.3mm}
\multiput(10,25)(0.16,0.12){25}{\line(1,0){0.16}}
\linethickness{0.3mm}
\put(14,28){\line(1,0){14}}
\linethickness{0.3mm}
\multiput(25,25)(0.12,0.12){25}{\line(1,0){0.12}}
\linethickness{0.3mm}
\put(28,13){\line(0,1){15}}
\linethickness{0.3mm}
\multiput(25,10)(0.12,0.12){25}{\line(1,0){0.12}}
\linethickness{0.1mm}
\put(14,13){\line(0,1){15}}
\linethickness{0.1mm}
\put(14,13){\line(1,0){14}}
\linethickness{0.1mm}
\multiput(10,10)(0.16,0.12){25}{\line(1,0){0.16}}
}


\put(60,0){
\linethickness{0.4mm}
\put(10,10){\line(0,1){20}}
\linethickness{0.4mm}
\put(10,30){\line(1,0){20}}
\linethickness{0.4mm}
\put(30,10){\line(0,1){20}}
\linethickness{0.4mm}
\put(10,10){\line(1,0){20}}
\linethickness{0.4mm}
\multiput(10,30)(0.18,0.12){167}{\line(1,0){0.18}}
\linethickness{0.4mm}
\multiput(30,30)(0.18,0.12){167}{\line(1,0){0.18}}
\linethickness{0.4mm}
\put(40,50){\line(1,0){20}}
\linethickness{0.4mm}
\put(60,30){\line(0,1){20}}
\linethickness{0.4mm}
\multiput(30,10)(0.18,0.12){167}{\line(1,0){0.18}}
\linethickness{0.1mm}
\multiput(10,10)(0.18,0.12){167}{\line(1,0){0.18}}
\linethickness{0.1mm}
\put(40,30){\line(0,1){20}}
\linethickness{0.1mm}
\put(40,30){\line(1,0){20}}
\put(23,50){\makebox(0,0)[cc]{$x_{\mathrm{con}}=1$}}
\put(54,15){\makebox(0,0)[cc]{$x_{\mathrm{opp}}=1$}}
\put(44,34){\makebox(0,0)[cc]{$O$}}
\put(30,5){\makebox(0,0)[cc]{$x_{\mathrm{end}}=1$}}

{\color{red}
\linethickness{0.3mm}
\put(10,25){\line(1,0){20}}
\linethickness{0.3mm}
\multiput(30,25)(0.18,0.12){167}{\line(1,0){0.18}}
\linethickness{0.3mm}
\put(10,25){\line(0,1){5}}
\linethickness{0.3mm}
\put(10,30){\line(1,0){20}}
\linethickness{0.3mm}
\put(30,25){\line(0,1){5}}
\linethickness{0.3mm}
\multiput(30,30)(0.18,0.12){167}{\line(1,0){0.18}}
\linethickness{0.3mm}
\put(60,45){\line(0,1){5}}
\linethickness{0.3mm}
\put(40,50){\line(1,0){20}}
\linethickness{0.3mm}
\multiput(10,30)(0.18,0.12){167}{\line(1,0){0.18}}
\linethickness{0.1mm}
\multiput(11,25)(0.17,0.12){167}{\line(1,0){0.17}}
\linethickness{0.05mm}
\put(40,45){\line(0,1){5}}
\linethickness{0.1mm}
\put(40,45){\line(1,0){20}}
}

}

\put(120,0){
\linethickness{0.4mm}
\put(10,10){\line(0,1){20}}
\linethickness{0.4mm}
\put(10,30){\line(1,0){20}}
\linethickness{0.4mm}
\put(30,10){\line(0,1){20}}
\linethickness{0.4mm}
\put(10,10){\line(1,0){20}}
\linethickness{0.4mm}
\multiput(10,30)(0.18,0.12){167}{\line(1,0){0.18}}
\linethickness{0.4mm}
\multiput(30,30)(0.18,0.12){167}{\line(1,0){0.18}}
\linethickness{0.4mm}
\put(40,50){\line(1,0){20}}
\linethickness{0.4mm}
\put(60,30){\line(0,1){20}}
\linethickness{0.4mm}
\multiput(30,10)(0.18,0.12){167}{\line(1,0){0.18}}
\linethickness{0.1mm}
\multiput(10,10)(0.18,0.12){167}{\line(1,0){0.18}}
\linethickness{0.1mm}
\put(40,30){\line(0,1){20}}
\linethickness{0.1mm}
\put(40,30){\line(1,0){20}}
\put(23,50){\makebox(0,0)[cc]{$x_{\mathrm{con}}=1$}}
\put(54,15){\makebox(0,0)[cc]{$x_{\mathrm{opp}}=1$}}
\put(44,34){\makebox(0,0)[cc]{$O$}}
\put(30,5){\makebox(0,0)[cc]{$x_{\mathrm{end}}=1$}}

{\color{red}
\linethickness{0.3mm}
\put(30,10){\line(0,1){20}}
\linethickness{0.3mm}
\multiput(30,30)(0.18,0.12){167}{\line(1,0){0.18}}
\linethickness{0.3mm}
\put(60,30){\line(0,1){20}}
\linethickness{0.3mm}
\multiput(30,10)(0.18,0.12){167}{\line(1,0){0.18}}
\linethickness{0.3mm}
\put(25,10){\line(1,0){5}}
\linethickness{0.3mm}
\put(25,10){\line(0,1){20}}
\linethickness{0.3mm}
\put(25,30){\line(1,0){5}}
\linethickness{0.3mm}
\multiput(25,30)(0.18,0.12){167}{\line(1,0){0.18}}
\linethickness{0.3mm}
\put(55,50){\line(1,0){5}}
\linethickness{0.1mm}
\put(55,30){\line(0,1){20}}
\linethickness{0.1mm}
\put(55,30){\line(1,0){5}}
\linethickness{0.1mm}
\multiput(25,10)(0.18,0.12){167}{\line(1,0){0.18}}
}

}

\put(180,0){
\linethickness{0.4mm}
\put(10,10){\line(0,1){20}}
\linethickness{0.4mm}
\put(10,30){\line(1,0){20}}
\linethickness{0.4mm}
\put(30,10){\line(0,1){20}}
\linethickness{0.4mm}
\put(10,10){\line(1,0){20}}
\linethickness{0.4mm}
\multiput(10,30)(0.18,0.12){167}{\line(1,0){0.18}}
\linethickness{0.4mm}
\multiput(30,30)(0.18,0.12){167}{\line(1,0){0.18}}
\linethickness{0.4mm}
\put(40,50){\line(1,0){20}}
\linethickness{0.4mm}
\put(60,30){\line(0,1){20}}
\linethickness{0.4mm}
\multiput(30,10)(0.18,0.12){167}{\line(1,0){0.18}}
\linethickness{0.1mm}
\multiput(10,10)(0.18,0.12){167}{\line(1,0){0.18}}
\linethickness{0.1mm}
\put(40,30){\line(0,1){20}}
\linethickness{0.1mm}
\put(40,30){\line(1,0){20}}
\put(23,50){\makebox(0,0)[cc]{$x_{\mathrm{con}}=1$}}
\put(54,15){\makebox(0,0)[cc]{$x_{\mathrm{opp}}=1$}}
\put(44,34){\makebox(0,0)[cc]{$O$}}
\put(30,5){\makebox(0,0)[cc]{$x_{\mathrm{end}}=1$}}

{\color{blue}
\linethickness{0.2mm}
\put(40,30){\line(0,1){15}}
\linethickness{0.2mm}
\put(40,45){\line(1,0){15}}
\linethickness{0.15mm}
\put(55,30){\line(0,1){15}}
\linethickness{0.2mm}
\put(40,30){\line(1,0){15}}
\linethickness{0.2mm}
\put(15,13){\line(0,1){14}}
\linethickness{0.2mm}
\multiput(15,27)(0.17,0.12){150}{\line(1,0){0.17}}
\linethickness{0.3mm}
\put(15,13){\line(1,0){12}}
\linethickness{0.2mm}
\put(27,13){\line(0,1){14}}
\linethickness{0.2mm}
\put(15,27){\line(1,0){12}}
\linethickness{0.3mm}
\multiput(27,13)(0.2,0.12){142}{\line(1,0){0.2}}
\linethickness{0.2mm}
\multiput(27,27)(0.19,0.12){150}{\line(1,0){0.19}}
}

}

\end{picture}
\setlength{\belowcaptionskip}{-15pt}
\caption{The subsets for neither, concern, opportunity and begin respectively}
\label{sepptri}
\end{center}
\end{figure}

In Figure~\ref{spaceofevid} we show the axes to plot the space of evidence, which is the unit cube, where the origin $O$ is the most distant vertex from our point of view. 
In Figure~\ref{sepptri} we highlight the subsets of evidence where we believe that the four outcomes are true. 

Thus, in the second diagram in Figure~\ref{sepptri} in the red subset we have points in the neighbourhood of the region where $x_{\mathrm{con}}=1$, i.e.,\ where we believe that we should be at the vertex \textit{concern}. 

In the third diagram in Figure~\ref{sepptri} in the red subset we have points in the neighbourhood of the region where $x_{\mathrm{opp}}=1$, i.e.,\ where we believe that we should be at the vertex \textit{opportunity}, and similarly for the leftmost diagram and the 
vertex \textit{opportunity}. The vertex \textit{begin} has a subset on the right given by a cube $[0,1-\epsilon)^3$ for some small $\epsilon>0$. 

\medskip
\noindent\textbf{Quantification: Belief and visualisation.}
To define the belief/visualisation function, we now construct a partition of unity for the cube $[0,1]^3$ (as in Definition~\ref{realpu} in the Appendix) for the above subsets. 
The subsets for \textit{opportunity} and \textit{concern} intersect as we can have an output along the edge linking those vertices. However, we do not allow the subset for \textit{end} intersect subsets for \textit{opportunity} and \textit{concern}. as those are not valid outputs on Figure~\ref{sectri}. This is because making discrete decisions from continuous data requires some amount of discontinuity -- made easier in our case as our data are actually rational numbers. Part of the choice of belief function is minimising the likelihood of discontinuity to make the system as transparent in real time (i.e.,\ as predictable in the short term) as possible. 

\subsection{Cyber architectures and ecosystems} \label{cyber_architectures}
  
In responding to a cyber attack on a component of a system concerns arise for the system architecture and its ecosystem, \textit{viz.} other components of the system, the external services the system depends upon, and the external systems it serves. Commonly, some of these dependencies may be unclear or unknown. Systems can be viewed at high level (simpler, hopefully transparent) and low level (complicated, likely opaque, due to lots of code and platform dependencies).

\medskip
\noindent \textbf{Scenario.}  
A problem is decisions made at a low level, e.g., using a buffer overflow to change the status of a block of code from non-executable to executable. This occurred in EternalBlue, the exploit behind the WannaCry ransomware cyber attack of 2017, which effected National Health Service (NHS) hospitals in the UK. One reason why some NHS systems were vulnerable to WannaCry was the problem of upgrading systems which have to be online continuously. The WannaCry worm bulk encrypted data on the affected machines, so systems which had reliable backups were less affected.
The WannaCry cyber attack was effectively halted by the activation of a built in kill-switch by registering an external web site, a method implemented by an individual about 7 hours after the attack began. Though quick enough to stop much more damage being caused by WannaCry, such a timescale may not be much use against a more malicious attack, e.g.,\ data theft, timed disruption of vital services, or physical damage. Such an attack would possibly use security flaws not used before. Human intervention takes time, and counter-measures may be difficult to circulate as some communication systems may have been affected or taken out by the attack. 

The first line of response is the isolation of critical systems as much as possible. This requires designs with inherent resistance to attack, and scenarios and exercises to prepare for response. 

As with a pandemic, an obvious line to take is compartmentalisation -- isolating things from each other to make spread difficult by not sharing resources. How do these observations relate to our concept of definition of a system of modes? We consider four general security features which can be described using modes.

\medskip
\noindent\textbf{Implementing updates during runtime -- shadow modes.}
Consider the idea that modes can be `shadowed' by duplicate modes, whose job is to have the same data and operations as the original mode. If the original mode needs to be taken offline (e.g., for upgrade) the shadow mode can be switched into its place. A shadow mode could also operate under a time delay or different security options from the original mode, making it less likely to be compromised in an attack than the original mode, but at the cost of not having quite the same data.

\begin{figure}[htbp]
\begin{center}
\unitlength 0.7 mm
\begin{picture}(110,35)(0,20)
\linethickness{0.3mm}
\put(0,40){\line(1,0){40}}
\linethickness{0.3mm}
\multiput(70,40)(0.24,-0.12){83}{\line(1,0){0.24}}
\linethickness{0.3mm}
\put(90,30){\line(0,1){20}}
\linethickness{0.3mm}
\multiput(70,40)(0.24,0.12){83}{\line(1,0){0.24}}
\linethickness{0.3mm}
\multiput(90,50)(0.24,-0.12){83}{\line(1,0){0.24}}
\linethickness{0.3mm}
\multiput(90,30)(0.24,0.12){83}{\line(1,0){0.24}}
\linethickness{0.1mm}
\multiput(79.38,37.5)(0.12,0.16){36}{\line(0,1){0.16}}
\linethickness{0.1mm}
\multiput(84.38,35.62)(0.12,0.18){31}{\line(0,1){0.18}}
\linethickness{0.1mm}
\multiput(92.5,33.75)(0.12,0.24){42}{\line(0,1){0.24}}
\linethickness{0.1mm}
\multiput(91.88,42.5)(0.12,0.23){16}{\line(0,1){0.23}}
\linethickness{0.1mm}
\multiput(97.5,35.62)(0.12,0.24){26}{\line(0,1){0.24}}
\linethickness{0.1mm}
\multiput(77.5,39.38)(0.12,0.12){16}{\line(1,0){0.12}}

\put(20,40){\makebox(0,0)[cc]{$\bullet$}}

\put(40,40){\makebox(0,0)[cc]{$\bullet$}}

\put(0,40){\makebox(0,0)[cc]{$\bullet$}}

\put(70,40){\makebox(0,0)[cc]{$\bullet$}}

\put(90,50){\makebox(0,0)[cc]{$\bullet$}}

\put(90,30){\makebox(0,0)[cc]{$\bullet$}}

\put(110,40){\makebox(0,0)[cc]{$\bullet$}}

\put(0,35){\makebox(0,0)[cc]{$\beta$}}

\put(20,35){\makebox(0,0)[cc]{$\alpha$}}

\put(40,35){\makebox(0,0)[cc]{$\gamma$}}

\put(70,35){\makebox(0,0)[cc]{$\beta$}}

\put(110,35){\makebox(0,0)[cc]{$\gamma$}}

\put(90,25){\makebox(0,0)[cc]{$\alpha$}}

\put(90,55){\makebox(0,0)[cc]{$\alpha'$}}

\end{picture}

\setlength{\belowcaptionskip}{-15pt}
\caption{Adding a shadow mode $\alpha'$}
\label{sepptrit}
\end{center}
\end{figure}
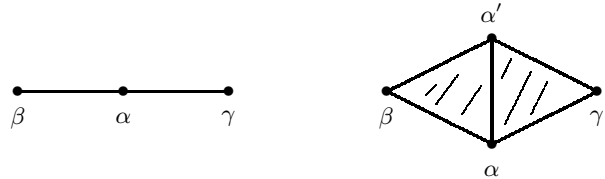

Symbolically, start with a collection of basic modes $\mathcal{M}=\{\alpha,\beta,\gamma,\dots\}$. We now add a `shadow' mode $\alpha'$ for $\alpha$, giving $\mathcal{M}'=\{\alpha,\alpha',\beta,\gamma,\dots\}$, as in Figure~\ref{sepptrit}. When the shadow is activated the effect is to copy all modes containing $\alpha$ -- in Figure~\ref{sepptrit} we have copies $\{\alpha'\}$, $\{\alpha',\beta\}$, $\{\alpha',\gamma\}$ of $\{\alpha\}$, $\{\alpha,\beta\}$, $\{\alpha,\gamma\}$ respectively. We also have modes $\{\alpha,\alpha'\}$, $\{\alpha,\alpha',\beta\}$, $\{\alpha,\alpha',\gamma\}$ which link the shadow and original modes. 

When the shadow is set up its first task for the linking modes (say for $\{\alpha,\alpha',\beta\}$)  is to copy the state and data for $\{\alpha,\beta\}$ to $\{\alpha',\beta\}$. It will continue to maintain the shadow copy $\{\alpha',\beta\}$ up to date, but all actual decisions will be taken by the dominant copy $\{\alpha,\beta\}$, the log for $\{\alpha',\beta\}$  will be kept for comparison to determine if there is any significant difference between the copies. (This could be as a result of a hardware problem or one of the copied being compromised.) At some point the roles of the shadow and dominant copy can be reversed, and then the new shadow copy $\alpha$ and can be shut down and updated or repaired.

\medskip
\noindent\textbf{Compartmentalisation of function.} Modes are assigned according to function, and are given only the tools (file access, external links, etc.) that are needed to perform that function. Regarding WannaCry, the options to bulk delete or encrypt data would simply not be available in most modes. It is not a matter of rewriting their code, the options to do that would not be on their allowed list of operations or they would not have the codes to access those options. They would not even know where to go to be able to access those codes. 


A malicious worm would likely have to be able to go through several modes and their isolation mechanisms before it was in a position to cause harm. But this movement at the high level of the modes is just what transparency and visualisation of the mode diagram is designed to detect, whether by a human or an AI supervisor who could then shut down the system or take other measures.

\medskip
\noindent\textbf{Isolation of modes.}
Different modes do not share software with each other. At the design stage methods or classes may be 
 inherited based on inclusion of modes -- i.e.,  mode $\{\alpha,\beta\}$ may inherit from $\{\alpha\}$.  However these will be compiled into distinct copies in memory, so there is no common memory at runtime. 
Thus, if one mode is corrupted the other modes will be unaffected unless a virus manages to propagate from mode to mode or send corrupted data. This forms a bottleneck in the infection process which can be examined in more detail.
When an alarm is raised communication between modes may be further restricted.

\medskip
\noindent\textbf{External monitoring of modes.}
As operations on any external agency (e.g., databases) are performed by oracle calls from the modes, the number of uses of such calls can be logged by a system not connected to the current operating mode, and an alarm raised if usage exceeds a prescribed amount.


\subsection{Critical incidents and the gold-silver-bronze command structures}\label{GSB-command}

Various scenarios are involved in preparing plans for responding to critical incidents, which should be able to generate simulations. In a multi-agency response to a critical incident streams of decisions need to be made by different commands drawing on different streams of information about an unfolding situation.  However, this can easily lead to confusion and information overload as hundreds of pieces of information, many inaccurate, flood in. Triaging, compiling and interpreting this information to formulate beliefs about the situation and to suggest a number of options are necessary tasks; how could these be supported  orpartially automated?  We examine the general form of the gold-silver-brone command structure, discussed in Section \ref{GSB}, though a toy example of a critical incident. These structures are likely to be broadly similar in many incidents.

\medskip
\noindent \textbf{Scenario.} Consider an event where a bomb has been detonated in a football stadium during a game.

Our first point of view is that of taking control of the incident. In general, after an emergency 999 call, one of the emergency services (police, ambulance, fire) will attend. But in our case, police will be on site and will radio their control to declare a `major incident', thus initiating the \textit{gold-silver-bronze command structure}.

In our scenario, all three services are on site.  Each of the services will set up their own silver command structures who will be in operational charge of the incident and will be located on or close to the incident. In our case, it is likely that a police inspector or superintendent would be at the match and would take this position. The fire and ambulance services should also have their silver commander on site, and preferably all the silver commanders would gather at the same place to allow free communication. A separate communications channel will likely be cleared for the incident.\footnote{If the military (other than explosive experts) become involved they may take overall charge.} 

The bronze commanders for each of the services will be in charge of various physical locations in the incident, and will report to their silver commander. The number and deployment of the bronze commanders will vary greatly and depend on the nature of the incident.

The gold command is the top-level strategic post, which establishes strategy for the combined operation, gathers and allocates resources and liaises with government, and other organisations and agencies.

 \begin{figure}[htbp]
\begin{center}
\unitlength 0.8 mm
\begin{picture}(130,57)(0,23)
\linethickness{0.3mm}
\multiput(12.5,47.5)(0.12,0.12){104}{\line(1,0){0.12}}
\linethickness{0.3mm}
\multiput(12.5,47.5)(0.28,-0.12){63}{\line(1,0){0.28}}
\linethickness{0.3mm}
\multiput(12.5,47.5)(0.44,0.12){63}{\line(1,0){0.44}}
\linethickness{0.3mm}
\multiput(25,60)(0.36,-0.12){42}{\line(1,0){0.36}}
\linethickness{0.3mm}
\multiput(30,40)(0.12,0.18){83}{\line(0,1){0.18}}
\linethickness{0.1mm}
\multiput(25,60)(0.12,-0.48){42}{\line(0,-1){0.48}}
\linethickness{0.3mm}
\multiput(25,60)(0.39,0.12){83}{\line(1,0){0.39}}
\linethickness{0.3mm}
\multiput(40,55)(0.36,0.12){83}{\line(1,0){0.36}}
\linethickness{0.3mm}
\multiput(30,40)(0.36,0.12){83}{\line(1,0){0.36}}
\linethickness{0.3mm}
\multiput(57.5,70)(0.3,-0.12){42}{\line(1,0){0.3}}
\linethickness{0.3mm}
\multiput(60,50)(0.12,0.18){83}{\line(0,1){0.18}}
\linethickness{0.1mm}
\multiput(57.5,70)(0.12,-0.95){21}{\line(0,-1){0.95}}
\linethickness{0.3mm}
\multiput(57.5,70)(1.19,-0.12){21}{\line(1,0){1.19}}
\linethickness{0.3mm}
\multiput(70,65)(0.6,0.12){21}{\line(1,0){0.6}}
\linethickness{0.3mm}
\multiput(60,50)(0.15,0.12){146}{\line(1,0){0.15}}

\put(0,5){
\linethickness{0.3mm}
\multiput(100,50)(0.12,0.12){125}{\line(1,0){0.12}}
\linethickness{0.3mm}
\multiput(115,65)(0.12,-0.18){83}{\line(0,-1){0.18}}
\linethickness{0.3mm}
\put(115,40){\line(0,1){25}}
\linethickness{0.3mm}
\multiput(115,40)(0.12,0.12){83}{\line(1,0){0.12}}
\linethickness{0.3mm}
\multiput(100,50)(0.18,-0.12){83}{\line(1,0){0.18}}
\linethickness{0.1mm}
\put(100,50){\line(1,0){25}}
}

\linethickness{0.3mm}
\multiput(27.5,65)(0.36,0.12){63}{\line(1,0){0.36}}
\put(50,72.5){\vector(3,1){0.12}}
\linethickness{0.3mm}
\multiput(40,37.5)(0.32,0.12){63}{\line(1,0){0.32}}
\put(40,37.5){\vector(-3,-1){0.12}}
\put(114,29){\makebox(0,0)[cc]{Au Ambulance}}

\put(70,29){\makebox(0,0)[cc]{Au Police}}

\put(81,40){\makebox(0,0)[cc]{Au Fire}}

\put(115,73){\makebox(0,0)[cc]{Politicians}}

\put(58,38){\makebox(0,0)[cc]{Stand down}}

\put(17,71){\makebox(0,0)[cc]{Major incident declared}}

\put(3,50){\makebox(0,0)[cc]{Incoming}}
\put(4,46){\makebox(0,0)[cc]{call}}

\put(43,52){\makebox(0,0)[cc]{Fire}}

\put(12.5,61){\makebox(0,0)[cc]{Ambulance}}

\put(30,37){\makebox(0,0)[cc]{Police}}

\put(68,73){\makebox(0,0)[cc]{Ag Ambulance}}

\put(57.5,65){\makebox(0,0)[cc]{Ag Fire}}

\put(69,47){\makebox(0,0)[cc]{Ag Police}}

\put(92,67){\makebox(0,0)[cc]{Military}}


\linethickness{0.3mm}
\multiput(90,40)(0.12,-0.15){83}{\line(0,-1){0.15}}
\linethickness{0.3mm}
\multiput(80,27.5)(0.12,0.15){83}{\line(0,1){0.15}}
\linethickness{0.3mm}
\put(80,27.5){\line(1,0){20}}
\linethickness{0.5mm}
\multiput(92.5,42.5)(0.12,0.19){54}{\line(0,1){0.19}}
\put(99,53){\vector(2,3){0.12}}
\put(90,32.5){\makebox(0,0)[cc]{Au}}

\put(96,55){\makebox(0,0)[cc]{Au}}


\put(132,55){\makebox(0,0)[cc]{Media}}

\put(115,42){\makebox(0,0)[cc]{Local Council}}

\end{picture}

\setlength{\belowcaptionskip}{-15pt}
\caption{The silver (Ag) and gold (Au) command structures}
\label{sepptrituyt}
\end{center}
\end{figure}
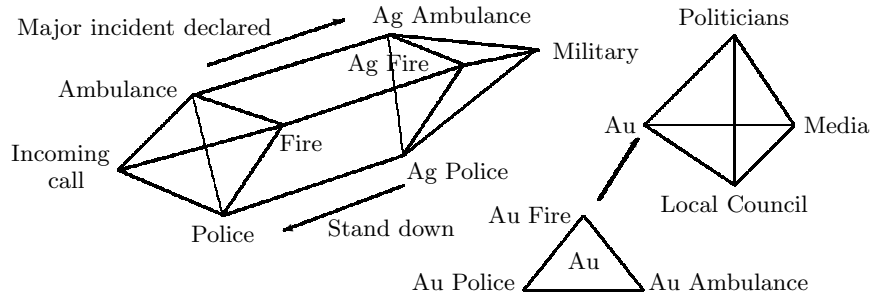

\medskip
\noindent \textbf{Modelling belief.}  Figure~\ref{sepptrituyt} shows the modes for setting up the silver command, building on the triage example in 
 Section~\ref{examples} and Figure~\ref{modpic}. The usual command structure is the police-fire-ambulance triangle, and this changes to silver command on a declaration of a major incident. We have shown an algebraic product of a triangle with an interval as a cylinder on the Figure, but in fact the modes are rather more complicated. This  would allow for delays in setting up some of the silver commands (real incidents are often chaotic). 
 
Figure~\ref{sepptrituyt} also shows the gold structure, based on coordination of the three services, and their links to the media, politicians (for resources) and local councils (for facilities such as shelter).

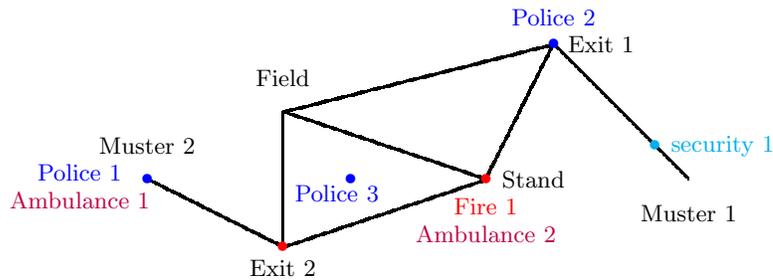
\begin{figure}[htbp]
\begin{center}
 \unitlength 0.9 mm
\begin{picture}(100,40)(0,33)
\linethickness{0.3mm}
\multiput(20,50)(0.24,-0.12){83}{\line(1,0){0.24}}
\linethickness{0.3mm}
\put(40,40){\line(0,1){20}}
\linethickness{0.3mm}
\multiput(40,60)(0.36,-0.12){83}{\line(1,0){0.36}}
\linethickness{0.3mm}
\multiput(40,40)(0.36,0.12){83}{\line(1,0){0.36}}
\linethickness{0.3mm}
\multiput(70,50)(0.12,0.24){83}{\line(0,1){0.24}}
\linethickness{0.3mm}
\multiput(40,60)(0.48,0.12){83}{\line(1,0){0.48}}
\linethickness{0.3mm}
\multiput(80,70)(0.12,-0.12){167}{\line(1,0){0.12}}

\put(95,55){\makebox(0,0)[cc]{ {\color{cyan}$\bullet$}  }}
\put(105,55){\makebox(0,0)[cc]{ {\color{cyan}security 1}  }}

\put(70,50){\makebox(0,0)[cc]{ {\color{red}$\bullet$} }}
\put(70,46){\makebox(0,0)[cc]{ {\color{red}Fire 1} }}
\put(70,42){\makebox(0,0)[cc]{ {\color{purple}Ambulance 2} }}

\put(50,50){\makebox(0,0)[cc]{ {\color{blue}$\bullet$}  }}
\put(48,48){\makebox(0,0)[cc]{ {\color{blue}Police 3}  }}

\put(20,50){\makebox(0,0)[cc]{ {\color{blue}$\bullet$} }}
\put(10,51){\makebox(0,0)[cc]{ {\color{blue}Police 1} }}
\put(10,47){\makebox(0,0)[cc]{ {\color{purple}Ambulance 1} }}


\put(80,70){\makebox(0,0)[cc]{ {\color{blue}$\bullet$} }}
\put(80,74){\makebox(0,0)[cc]{ {\color{blue}Police 2} }}

\put(40,40){\makebox(0,0)[cc]{ {\color{red}$\bullet$} }}


\put(77,50){\makebox(0,0)[cc]{Stand}}

\put(87,70){\makebox(0,0)[cc]{Exit 1}}

\put(40,65){\makebox(0,0)[cc]{Field}}

\put(40,37){\makebox(0,0)[cc]{Exit 2}}

\put(100,45){\makebox(0,0)[cc]{Muster 1}}

\put(20,55){\makebox(0,0)[cc]{Muster 2}}

\end{picture}

\setlength{\belowcaptionskip}{-15pt}
\caption{Bronze command and the geographical layout of teams}
\label{bronze}
\end{center}
\end{figure}

In principle, one bronze commander is assigned to each physical area of the site and is in personal command of the teams there. Figure~\ref{bronze} depicts the geographical layout of the areas in the stadium scenario, with the positions of teams indicated (the local stadium security coming under police command). Each area should have a bronze commander, who reports to the appropriate silver commander, and then communicates with the teams in the area. 


\section{Concluding remarks} \label{Potential}

\subsection{Reflections and next steps}

Given the questions posed in the Introduction, a general aim is to develop mathematical tools for thinking about current and new systems in complex human contexts, described by security scenarios.
Our modelling methods for systems, based on notions of mode and mode transition, seem to have some advantages in terms of security and resilience, e.g.,

\smallskip
\noindent \textit{Concurrency}: Several modes can be designed to work collaboratively on the same problems.

\noindent \textit{Independence}: The independent modes are designed so as to avoid, or at least contain, potential damage, and to facilitate auditing.

\noindent \textit{Continuity}:  Changes can be made to modes separately allowing updates without system downtime.  If a mode is behaving suspiciously, this should be recognised by the other modes, and the errant mode could be suspended. If one mode is down, others need not be compromised. 

\noindent \textit{Transparency}: The geometric visualisation supports analysis, transparency and explication.

\smallskip
As automation expands into new areas of life through AI, transparency is particularly important and timely.

In terms of next steps, there is much to explore in the types of examples discussed in this paper; for example, triage is a hugely important topic, one associated with machine learning.  Only though case studies, including realistic applications and past events, can we establish the evidence and extent the \textit{prima facie} advantages -- such as concurrency, independence, continuity, and transparency -- actually obtain.  By drilling down into the internals of the modes, and the mathematics of their geometries and belief functions, we can develop a general theory of modes and mode transitions. In due course, we will need to develop specifications for software tools to explore and strengthen the methods.


\subsection{Theorising scenarios}\label{conceptualsing_scenarios}
Setting aside dictionary definitions, the term scenario suggests an idealisation that postulates a context and dynamic events that are abstracted  from the real world.\footnote{Interestingly, the 1982 Supplement to the Oxford English Dictionary remarked ``The over-use of this word in various loose senses has attracted frequent hostile comment''.}  The term security scenario suggests an account or synopsis of a possible unfolding of events and courses of reaction in which local or national security is implicated.
Such idealisations are needed for planning responses to security problems; they also shape evaluations, feedback and investigations of historical incidents. Indeed, our understanding of past security problems play an essential role in  preparations for future security problems. The same is true of history in military planning and training, of course.

It is plausible to argue that, ultimately, scenarios are both the raison d'\^{e}tre for systems as well as formidable tests for their evaluation - their nemesis in security terms.
So we might ask: 

\textit{What is the essential structure of a scenario? Can we develop a rigorous framework within which systems and scenarios can be compared? Can such a framework be formalised?}

Over the past two decades, scenarios have also been used as a tool to deal with the complexities and uncertainties associated with global issues, such as climate change, food security, and land use. This allows the simulation of different environments, approaches and outcomes, as well as delivering a multitude of perspectives on potential future developments. Thus, scenario-based research has been frequently conducted by all sorts of professionals. To give a sample of fields: economics and business studies (e.g., \cite{Ozoran_etal2023}); political science and international relations (e.g., \cite{SusHadeed2020}); environmental science and climate studies (e.g., \cite{Flynnetal2018}); public health and epidemiology (e.g., \cite{Anderson_etal2019}); urban planning and architecture (e.g., \cite{Eilouti2018}); psychology and sociology (e.g., \cite{Bekkers2010}); military war-gaming (e.g., \cite{Augier_etal2018}); and disaster management and emergency planning (e.g., \cite{Li&Wang2022}).  One commonly mentioned general description is: `plausible and often simplified descriptions of how the future may develop based on a coherent and internally consistent set of assumptions on key driving forces and relationships' (\cite{Carpenter_etal2005}). The set of assumptions are key to the construction of scenarios and their varied application.

Looking ahead at what may count as threats and risks (say, over a decade) suggest to us that new tools and methodologies for thinking about security scenarios, idealised and actual, should be useful and are, in fact, needed. Scenarios play an essential role in creating and judging security systems and in assessing risk more broadly.

Making a stocktake of technologies that we know currently and which may come to some sort of maturity in the coming decade is an interesting exercise (e.g., \cite{Diana2016}). Every new technology can generate new surprises for security, broadly conceived. Security surprises can also come from ill-understood functional interdependencies of: current systems, or mutual dependent sets of users, or underlying, shared computing infrastructures. Most importantly, we note that \textit{our surprise may be proportional to our \textit{ignorance} of these factors and their effects}. If the critical technologies of the next decade are unknown then security surprises are classic `unknown unknowns'.  Thus, to avoid, or at least partially prepare, for surprises we may turn to scenarios.

We believe it sensible to seek new theories about security scenarios that abstract from lived and imagined experiences and, specifically, seek rigorous, systematic and useable methodologies, some mathematical models, and software tools. In particular, they must make explicit the various roles of data and automation play in decisions by the systems involved. 

%
%
%
%
%
%
%
%
%
%
%

\bibliographystyle{compj}

\newpage

\section{Appendix: The mathematics}
\label{app}

\subsection{Logical and geometric methods} \label{math}
In Definition~\ref{mode_working definition} a mode is described in terms of subsets of the states and objectives of a system. The mathematics of how to deal with a set built as a union of subsets has been known for a long time -- a simplicial complex \cite{LeeManif}.

\begin{figure}[htbp]
\begin{center}
\unitlength 0.5 mm
\begin{picture}(110,60)(0,30)
\linethickness{0.3mm}
\put(30,74.75){\line(0,1){0.51}}
\multiput(29.97,74.24)(0.03,0.51){1}{\line(0,1){0.51}}
\multiput(29.92,73.74)(0.05,0.5){1}{\line(0,1){0.5}}
\multiput(29.84,73.24)(0.08,0.5){1}{\line(0,1){0.5}}
\multiput(29.74,72.74)(0.1,0.5){1}{\line(0,1){0.5}}
\multiput(29.61,72.25)(0.13,0.49){1}{\line(0,1){0.49}}
\multiput(29.46,71.77)(0.15,0.48){1}{\line(0,1){0.48}}
\multiput(29.29,71.29)(0.18,0.48){1}{\line(0,1){0.48}}
\multiput(29.09,70.82)(0.1,0.23){2}{\line(0,1){0.23}}
\multiput(28.86,70.37)(0.11,0.23){2}{\line(0,1){0.23}}
\multiput(28.62,69.93)(0.12,0.22){2}{\line(0,1){0.22}}
\multiput(28.35,69.5)(0.13,0.21){2}{\line(0,1){0.21}}
\multiput(28.06,69.08)(0.14,0.21){2}{\line(0,1){0.21}}
\multiput(27.75,68.68)(0.1,0.13){3}{\line(0,1){0.13}}
\multiput(27.42,68.3)(0.11,0.13){3}{\line(0,1){0.13}}
\multiput(27.07,67.93)(0.12,0.12){3}{\line(0,1){0.12}}
\multiput(26.7,67.58)(0.12,0.12){3}{\line(1,0){0.12}}
\multiput(26.32,67.25)(0.13,0.11){3}{\line(1,0){0.13}}
\multiput(25.92,66.94)(0.13,0.1){3}{\line(1,0){0.13}}
\multiput(25.5,66.65)(0.21,0.14){2}{\line(1,0){0.21}}
\multiput(25.07,66.38)(0.21,0.13){2}{\line(1,0){0.21}}
\multiput(24.63,66.14)(0.22,0.12){2}{\line(1,0){0.22}}
\multiput(24.18,65.91)(0.23,0.11){2}{\line(1,0){0.23}}
\multiput(23.71,65.71)(0.23,0.1){2}{\line(1,0){0.23}}
\multiput(23.23,65.54)(0.48,0.18){1}{\line(1,0){0.48}}
\multiput(22.75,65.39)(0.48,0.15){1}{\line(1,0){0.48}}
\multiput(22.26,65.26)(0.49,0.13){1}{\line(1,0){0.49}}
\multiput(21.76,65.16)(0.5,0.1){1}{\line(1,0){0.5}}
\multiput(21.26,65.08)(0.5,0.08){1}{\line(1,0){0.5}}
\multiput(20.76,65.03)(0.5,0.05){1}{\line(1,0){0.5}}
\multiput(20.25,65)(0.51,0.03){1}{\line(1,0){0.51}}
\put(19.75,65){\line(1,0){0.51}}
\multiput(19.24,65.03)(0.51,-0.03){1}{\line(1,0){0.51}}
\multiput(18.74,65.08)(0.5,-0.05){1}{\line(1,0){0.5}}
\multiput(18.24,65.16)(0.5,-0.08){1}{\line(1,0){0.5}}
\multiput(17.74,65.26)(0.5,-0.1){1}{\line(1,0){0.5}}
\multiput(17.25,65.39)(0.49,-0.13){1}{\line(1,0){0.49}}
\multiput(16.77,65.54)(0.48,-0.15){1}{\line(1,0){0.48}}
\multiput(16.29,65.71)(0.48,-0.18){1}{\line(1,0){0.48}}
\multiput(15.82,65.91)(0.23,-0.1){2}{\line(1,0){0.23}}
\multiput(15.37,66.14)(0.23,-0.11){2}{\line(1,0){0.23}}
\multiput(14.93,66.38)(0.22,-0.12){2}{\line(1,0){0.22}}
\multiput(14.5,66.65)(0.21,-0.13){2}{\line(1,0){0.21}}
\multiput(14.08,66.94)(0.21,-0.14){2}{\line(1,0){0.21}}
\multiput(13.68,67.25)(0.13,-0.1){3}{\line(1,0){0.13}}
\multiput(13.3,67.58)(0.13,-0.11){3}{\line(1,0){0.13}}
\multiput(12.93,67.93)(0.12,-0.12){3}{\line(1,0){0.12}}
\multiput(12.58,68.3)(0.12,-0.12){3}{\line(0,-1){0.12}}
\multiput(12.25,68.68)(0.11,-0.13){3}{\line(0,-1){0.13}}
\multiput(11.94,69.08)(0.1,-0.13){3}{\line(0,-1){0.13}}
\multiput(11.65,69.5)(0.14,-0.21){2}{\line(0,-1){0.21}}
\multiput(11.38,69.93)(0.13,-0.21){2}{\line(0,-1){0.21}}
\multiput(11.14,70.37)(0.12,-0.22){2}{\line(0,-1){0.22}}
\multiput(10.91,70.82)(0.11,-0.23){2}{\line(0,-1){0.23}}
\multiput(10.71,71.29)(0.1,-0.23){2}{\line(0,-1){0.23}}
\multiput(10.54,71.77)(0.18,-0.48){1}{\line(0,-1){0.48}}
\multiput(10.39,72.25)(0.15,-0.48){1}{\line(0,-1){0.48}}
\multiput(10.26,72.74)(0.13,-0.49){1}{\line(0,-1){0.49}}
\multiput(10.16,73.24)(0.1,-0.5){1}{\line(0,-1){0.5}}
\multiput(10.08,73.74)(0.08,-0.5){1}{\line(0,-1){0.5}}
\multiput(10.03,74.24)(0.05,-0.5){1}{\line(0,-1){0.5}}
\multiput(10,74.75)(0.03,-0.51){1}{\line(0,-1){0.51}}
\put(10,74.75){\line(0,1){0.51}}
\multiput(10,75.25)(0.03,0.51){1}{\line(0,1){0.51}}
\multiput(10.03,75.76)(0.05,0.5){1}{\line(0,1){0.5}}
\multiput(10.08,76.26)(0.08,0.5){1}{\line(0,1){0.5}}
\multiput(10.16,76.76)(0.1,0.5){1}{\line(0,1){0.5}}
\multiput(10.26,77.26)(0.13,0.49){1}{\line(0,1){0.49}}
\multiput(10.39,77.75)(0.15,0.48){1}{\line(0,1){0.48}}
\multiput(10.54,78.23)(0.18,0.48){1}{\line(0,1){0.48}}
\multiput(10.71,78.71)(0.1,0.23){2}{\line(0,1){0.23}}
\multiput(10.91,79.18)(0.11,0.23){2}{\line(0,1){0.23}}
\multiput(11.14,79.63)(0.12,0.22){2}{\line(0,1){0.22}}
\multiput(11.38,80.07)(0.13,0.21){2}{\line(0,1){0.21}}
\multiput(11.65,80.5)(0.14,0.21){2}{\line(0,1){0.21}}
\multiput(11.94,80.92)(0.1,0.13){3}{\line(0,1){0.13}}
\multiput(12.25,81.32)(0.11,0.13){3}{\line(0,1){0.13}}
\multiput(12.58,81.7)(0.12,0.12){3}{\line(0,1){0.12}}
\multiput(12.93,82.07)(0.12,0.12){3}{\line(1,0){0.12}}
\multiput(13.3,82.42)(0.13,0.11){3}{\line(1,0){0.13}}
\multiput(13.68,82.75)(0.13,0.1){3}{\line(1,0){0.13}}
\multiput(14.08,83.06)(0.21,0.14){2}{\line(1,0){0.21}}
\multiput(14.5,83.35)(0.21,0.13){2}{\line(1,0){0.21}}
\multiput(14.93,83.62)(0.22,0.12){2}{\line(1,0){0.22}}
\multiput(15.37,83.86)(0.23,0.11){2}{\line(1,0){0.23}}
\multiput(15.82,84.09)(0.23,0.1){2}{\line(1,0){0.23}}
\multiput(16.29,84.29)(0.48,0.18){1}{\line(1,0){0.48}}
\multiput(16.77,84.46)(0.48,0.15){1}{\line(1,0){0.48}}
\multiput(17.25,84.61)(0.49,0.13){1}{\line(1,0){0.49}}
\multiput(17.74,84.74)(0.5,0.1){1}{\line(1,0){0.5}}
\multiput(18.24,84.84)(0.5,0.08){1}{\line(1,0){0.5}}
\multiput(18.74,84.92)(0.5,0.05){1}{\line(1,0){0.5}}
\multiput(19.24,84.97)(0.51,0.03){1}{\line(1,0){0.51}}
\put(19.75,85){\line(1,0){0.51}}
\multiput(20.25,85)(0.51,-0.03){1}{\line(1,0){0.51}}
\multiput(20.76,84.97)(0.5,-0.05){1}{\line(1,0){0.5}}
\multiput(21.26,84.92)(0.5,-0.08){1}{\line(1,0){0.5}}
\multiput(21.76,84.84)(0.5,-0.1){1}{\line(1,0){0.5}}
\multiput(22.26,84.74)(0.49,-0.13){1}{\line(1,0){0.49}}
\multiput(22.75,84.61)(0.48,-0.15){1}{\line(1,0){0.48}}
\multiput(23.23,84.46)(0.48,-0.18){1}{\line(1,0){0.48}}
\multiput(23.71,84.29)(0.23,-0.1){2}{\line(1,0){0.23}}
\multiput(24.18,84.09)(0.23,-0.11){2}{\line(1,0){0.23}}
\multiput(24.63,83.86)(0.22,-0.12){2}{\line(1,0){0.22}}
\multiput(25.07,83.62)(0.21,-0.13){2}{\line(1,0){0.21}}
\multiput(25.5,83.35)(0.21,-0.14){2}{\line(1,0){0.21}}
\multiput(25.92,83.06)(0.13,-0.1){3}{\line(1,0){0.13}}
\multiput(26.32,82.75)(0.13,-0.11){3}{\line(1,0){0.13}}
\multiput(26.7,82.42)(0.12,-0.12){3}{\line(1,0){0.12}}
\multiput(27.07,82.07)(0.12,-0.12){3}{\line(0,-1){0.12}}
\multiput(27.42,81.7)(0.11,-0.13){3}{\line(0,-1){0.13}}
\multiput(27.75,81.32)(0.1,-0.13){3}{\line(0,-1){0.13}}
\multiput(28.06,80.92)(0.14,-0.21){2}{\line(0,-1){0.21}}
\multiput(28.35,80.5)(0.13,-0.21){2}{\line(0,-1){0.21}}
\multiput(28.62,80.07)(0.12,-0.22){2}{\line(0,-1){0.22}}
\multiput(28.86,79.63)(0.11,-0.23){2}{\line(0,-1){0.23}}
\multiput(29.09,79.18)(0.1,-0.23){2}{\line(0,-1){0.23}}
\multiput(29.29,78.71)(0.18,-0.48){1}{\line(0,-1){0.48}}
\multiput(29.46,78.23)(0.15,-0.48){1}{\line(0,-1){0.48}}
\multiput(29.61,77.75)(0.13,-0.49){1}{\line(0,-1){0.49}}
\multiput(29.74,77.26)(0.1,-0.5){1}{\line(0,-1){0.5}}
\multiput(29.84,76.76)(0.08,-0.5){1}{\line(0,-1){0.5}}
\multiput(29.92,76.26)(0.05,-0.5){1}{\line(0,-1){0.5}}
\multiput(29.97,75.76)(0.03,-0.51){1}{\line(0,-1){0.51}}

\linethickness{0.3mm}
\put(40,72.25){\line(0,1){0.5}}
\multiput(39.99,71.75)(0.01,0.5){1}{\line(0,1){0.5}}
\multiput(39.98,71.25)(0.01,0.5){1}{\line(0,1){0.5}}
\multiput(39.96,70.75)(0.02,0.5){1}{\line(0,1){0.5}}
\multiput(39.94,70.26)(0.02,0.5){1}{\line(0,1){0.5}}
\multiput(39.91,69.76)(0.03,0.49){1}{\line(0,1){0.49}}
\multiput(39.87,69.27)(0.04,0.49){1}{\line(0,1){0.49}}
\multiput(39.83,68.78)(0.04,0.49){1}{\line(0,1){0.49}}
\multiput(39.78,68.29)(0.05,0.49){1}{\line(0,1){0.49}}
\multiput(39.73,67.81)(0.05,0.48){1}{\line(0,1){0.48}}
\multiput(39.67,67.33)(0.06,0.48){1}{\line(0,1){0.48}}
\multiput(39.6,66.86)(0.07,0.48){1}{\line(0,1){0.48}}
\multiput(39.53,66.38)(0.07,0.47){1}{\line(0,1){0.47}}
\multiput(39.45,65.92)(0.08,0.47){1}{\line(0,1){0.47}}
\multiput(39.37,65.46)(0.08,0.46){1}{\line(0,1){0.46}}
\multiput(39.28,65)(0.09,0.45){1}{\line(0,1){0.45}}
\multiput(39.18,64.56)(0.09,0.45){1}{\line(0,1){0.45}}
\multiput(39.08,64.11)(0.1,0.44){1}{\line(0,1){0.44}}
\multiput(38.98,63.68)(0.11,0.44){1}{\line(0,1){0.44}}
\multiput(38.87,63.25)(0.11,0.43){1}{\line(0,1){0.43}}
\multiput(38.75,62.83)(0.12,0.42){1}{\line(0,1){0.42}}
\multiput(38.63,62.42)(0.12,0.41){1}{\line(0,1){0.41}}
\multiput(38.5,62.01)(0.13,0.4){1}{\line(0,1){0.4}}
\multiput(38.37,61.62)(0.13,0.4){1}{\line(0,1){0.4}}
\multiput(38.24,61.23)(0.14,0.39){1}{\line(0,1){0.39}}
\multiput(38.1,60.85)(0.14,0.38){1}{\line(0,1){0.38}}
\multiput(37.95,60.48)(0.14,0.37){1}{\line(0,1){0.37}}
\multiput(37.8,60.13)(0.15,0.36){1}{\line(0,1){0.36}}
\multiput(37.65,59.78)(0.15,0.35){1}{\line(0,1){0.35}}
\multiput(37.49,59.44)(0.16,0.34){1}{\line(0,1){0.34}}
\multiput(37.33,59.11)(0.16,0.33){1}{\line(0,1){0.33}}
\multiput(37.16,58.8)(0.17,0.32){1}{\line(0,1){0.32}}
\multiput(36.99,58.49)(0.17,0.31){1}{\line(0,1){0.31}}
\multiput(36.82,58.2)(0.17,0.29){1}{\line(0,1){0.29}}
\multiput(36.64,57.91)(0.18,0.28){1}{\line(0,1){0.28}}
\multiput(36.46,57.64)(0.09,0.14){2}{\line(0,1){0.14}}
\multiput(36.28,57.39)(0.09,0.13){2}{\line(0,1){0.13}}
\multiput(36.09,57.14)(0.09,0.12){2}{\line(0,1){0.12}}
\multiput(35.9,56.91)(0.09,0.12){2}{\line(0,1){0.12}}
\multiput(35.71,56.69)(0.1,0.11){2}{\line(0,1){0.11}}
\multiput(35.52,56.48)(0.1,0.1){2}{\line(0,1){0.1}}
\multiput(35.32,56.28)(0.1,0.1){2}{\line(1,0){0.1}}
\multiput(35.12,56.1)(0.1,0.09){2}{\line(1,0){0.1}}
\multiput(34.92,55.94)(0.2,0.17){1}{\line(1,0){0.2}}
\multiput(34.72,55.78)(0.2,0.15){1}{\line(1,0){0.2}}
\multiput(34.51,55.64)(0.21,0.14){1}{\line(1,0){0.21}}
\multiput(34.3,55.51)(0.21,0.13){1}{\line(1,0){0.21}}
\multiput(34.09,55.4)(0.21,0.11){1}{\line(1,0){0.21}}
\multiput(33.88,55.3)(0.21,0.1){1}{\line(1,0){0.21}}
\multiput(33.67,55.22)(0.21,0.09){1}{\line(1,0){0.21}}
\multiput(33.46,55.14)(0.21,0.07){1}{\line(1,0){0.21}}
\multiput(33.25,55.09)(0.21,0.06){1}{\line(1,0){0.21}}
\multiput(33.04,55.04)(0.21,0.04){1}{\line(1,0){0.21}}
\multiput(32.82,55.02)(0.21,0.03){1}{\line(1,0){0.21}}
\multiput(32.61,55)(0.21,0.01){1}{\line(1,0){0.21}}
\put(32.39,55){\line(1,0){0.21}}
\multiput(32.18,55.02)(0.21,-0.01){1}{\line(1,0){0.21}}
\multiput(31.96,55.04)(0.21,-0.03){1}{\line(1,0){0.21}}
\multiput(31.75,55.09)(0.21,-0.04){1}{\line(1,0){0.21}}
\multiput(31.54,55.14)(0.21,-0.06){1}{\line(1,0){0.21}}
\multiput(31.33,55.22)(0.21,-0.07){1}{\line(1,0){0.21}}
\multiput(31.12,55.3)(0.21,-0.09){1}{\line(1,0){0.21}}
\multiput(30.91,55.4)(0.21,-0.1){1}{\line(1,0){0.21}}
\multiput(30.7,55.51)(0.21,-0.11){1}{\line(1,0){0.21}}
\multiput(30.49,55.64)(0.21,-0.13){1}{\line(1,0){0.21}}
\multiput(30.28,55.78)(0.21,-0.14){1}{\line(1,0){0.21}}
\multiput(30.08,55.94)(0.2,-0.15){1}{\line(1,0){0.2}}
\multiput(29.88,56.1)(0.2,-0.17){1}{\line(1,0){0.2}}
\multiput(29.68,56.28)(0.1,-0.09){2}{\line(1,0){0.1}}
\multiput(29.48,56.48)(0.1,-0.1){2}{\line(1,0){0.1}}
\multiput(29.29,56.69)(0.1,-0.1){2}{\line(0,-1){0.1}}
\multiput(29.1,56.91)(0.1,-0.11){2}{\line(0,-1){0.11}}
\multiput(28.91,57.14)(0.09,-0.12){2}{\line(0,-1){0.12}}
\multiput(28.72,57.39)(0.09,-0.12){2}{\line(0,-1){0.12}}
\multiput(28.54,57.64)(0.09,-0.13){2}{\line(0,-1){0.13}}
\multiput(28.36,57.91)(0.09,-0.14){2}{\line(0,-1){0.14}}
\multiput(28.18,58.2)(0.18,-0.28){1}{\line(0,-1){0.28}}
\multiput(28.01,58.49)(0.17,-0.29){1}{\line(0,-1){0.29}}
\multiput(27.84,58.8)(0.17,-0.31){1}{\line(0,-1){0.31}}
\multiput(27.67,59.11)(0.17,-0.32){1}{\line(0,-1){0.32}}
\multiput(27.51,59.44)(0.16,-0.33){1}{\line(0,-1){0.33}}
\multiput(27.35,59.78)(0.16,-0.34){1}{\line(0,-1){0.34}}
\multiput(27.2,60.13)(0.15,-0.35){1}{\line(0,-1){0.35}}
\multiput(27.05,60.48)(0.15,-0.36){1}{\line(0,-1){0.36}}
\multiput(26.9,60.85)(0.14,-0.37){1}{\line(0,-1){0.37}}
\multiput(26.76,61.23)(0.14,-0.38){1}{\line(0,-1){0.38}}
\multiput(26.63,61.62)(0.14,-0.39){1}{\line(0,-1){0.39}}
\multiput(26.5,62.01)(0.13,-0.4){1}{\line(0,-1){0.4}}
\multiput(26.37,62.42)(0.13,-0.4){1}{\line(0,-1){0.4}}
\multiput(26.25,62.83)(0.12,-0.41){1}{\line(0,-1){0.41}}
\multiput(26.13,63.25)(0.12,-0.42){1}{\line(0,-1){0.42}}
\multiput(26.02,63.68)(0.11,-0.43){1}{\line(0,-1){0.43}}
\multiput(25.92,64.11)(0.11,-0.44){1}{\line(0,-1){0.44}}
\multiput(25.82,64.56)(0.1,-0.44){1}{\line(0,-1){0.44}}
\multiput(25.72,65)(0.09,-0.45){1}{\line(0,-1){0.45}}
\multiput(25.63,65.46)(0.09,-0.45){1}{\line(0,-1){0.45}}
\multiput(25.55,65.92)(0.08,-0.46){1}{\line(0,-1){0.46}}
\multiput(25.47,66.38)(0.08,-0.47){1}{\line(0,-1){0.47}}
\multiput(25.4,66.86)(0.07,-0.47){1}{\line(0,-1){0.47}}
\multiput(25.33,67.33)(0.07,-0.48){1}{\line(0,-1){0.48}}
\multiput(25.27,67.81)(0.06,-0.48){1}{\line(0,-1){0.48}}
\multiput(25.22,68.29)(0.05,-0.48){1}{\line(0,-1){0.48}}
\multiput(25.17,68.78)(0.05,-0.49){1}{\line(0,-1){0.49}}
\multiput(25.13,69.27)(0.04,-0.49){1}{\line(0,-1){0.49}}
\multiput(25.09,69.76)(0.04,-0.49){1}{\line(0,-1){0.49}}
\multiput(25.06,70.26)(0.03,-0.49){1}{\line(0,-1){0.49}}
\multiput(25.04,70.75)(0.02,-0.5){1}{\line(0,-1){0.5}}
\multiput(25.02,71.25)(0.02,-0.5){1}{\line(0,-1){0.5}}
\multiput(25.01,71.75)(0.01,-0.5){1}{\line(0,-1){0.5}}
\multiput(25,72.25)(0.01,-0.5){1}{\line(0,-1){0.5}}
\put(25,72.25){\line(0,1){0.5}}
\multiput(25,72.75)(0.01,0.5){1}{\line(0,1){0.5}}
\multiput(25.01,73.25)(0.01,0.5){1}{\line(0,1){0.5}}
\multiput(25.02,73.75)(0.02,0.5){1}{\line(0,1){0.5}}
\multiput(25.04,74.25)(0.02,0.5){1}{\line(0,1){0.5}}
\multiput(25.06,74.74)(0.03,0.49){1}{\line(0,1){0.49}}
\multiput(25.09,75.24)(0.04,0.49){1}{\line(0,1){0.49}}
\multiput(25.13,75.73)(0.04,0.49){1}{\line(0,1){0.49}}
\multiput(25.17,76.22)(0.05,0.49){1}{\line(0,1){0.49}}
\multiput(25.22,76.71)(0.05,0.48){1}{\line(0,1){0.48}}
\multiput(25.27,77.19)(0.06,0.48){1}{\line(0,1){0.48}}
\multiput(25.33,77.67)(0.07,0.48){1}{\line(0,1){0.48}}
\multiput(25.4,78.14)(0.07,0.47){1}{\line(0,1){0.47}}
\multiput(25.47,78.62)(0.08,0.47){1}{\line(0,1){0.47}}
\multiput(25.55,79.08)(0.08,0.46){1}{\line(0,1){0.46}}
\multiput(25.63,79.54)(0.09,0.45){1}{\line(0,1){0.45}}
\multiput(25.72,80)(0.09,0.45){1}{\line(0,1){0.45}}
\multiput(25.82,80.44)(0.1,0.44){1}{\line(0,1){0.44}}
\multiput(25.92,80.89)(0.11,0.44){1}{\line(0,1){0.44}}
\multiput(26.02,81.32)(0.11,0.43){1}{\line(0,1){0.43}}
\multiput(26.13,81.75)(0.12,0.42){1}{\line(0,1){0.42}}
\multiput(26.25,82.17)(0.12,0.41){1}{\line(0,1){0.41}}
\multiput(26.37,82.58)(0.13,0.4){1}{\line(0,1){0.4}}
\multiput(26.5,82.99)(0.13,0.4){1}{\line(0,1){0.4}}
\multiput(26.63,83.38)(0.14,0.39){1}{\line(0,1){0.39}}
\multiput(26.76,83.77)(0.14,0.38){1}{\line(0,1){0.38}}
\multiput(26.9,84.15)(0.14,0.37){1}{\line(0,1){0.37}}
\multiput(27.05,84.52)(0.15,0.36){1}{\line(0,1){0.36}}
\multiput(27.2,84.87)(0.15,0.35){1}{\line(0,1){0.35}}
\multiput(27.35,85.22)(0.16,0.34){1}{\line(0,1){0.34}}
\multiput(27.51,85.56)(0.16,0.33){1}{\line(0,1){0.33}}
\multiput(27.67,85.89)(0.17,0.32){1}{\line(0,1){0.32}}
\multiput(27.84,86.2)(0.17,0.31){1}{\line(0,1){0.31}}
\multiput(28.01,86.51)(0.17,0.29){1}{\line(0,1){0.29}}
\multiput(28.18,86.8)(0.18,0.28){1}{\line(0,1){0.28}}
\multiput(28.36,87.09)(0.09,0.14){2}{\line(0,1){0.14}}
\multiput(28.54,87.36)(0.09,0.13){2}{\line(0,1){0.13}}
\multiput(28.72,87.61)(0.09,0.12){2}{\line(0,1){0.12}}
\multiput(28.91,87.86)(0.09,0.12){2}{\line(0,1){0.12}}
\multiput(29.1,88.09)(0.1,0.11){2}{\line(0,1){0.11}}
\multiput(29.29,88.31)(0.1,0.1){2}{\line(0,1){0.1}}
\multiput(29.48,88.52)(0.1,0.1){2}{\line(1,0){0.1}}
\multiput(29.68,88.72)(0.1,0.09){2}{\line(1,0){0.1}}
\multiput(29.88,88.9)(0.2,0.17){1}{\line(1,0){0.2}}
\multiput(30.08,89.06)(0.2,0.15){1}{\line(1,0){0.2}}
\multiput(30.28,89.22)(0.21,0.14){1}{\line(1,0){0.21}}
\multiput(30.49,89.36)(0.21,0.13){1}{\line(1,0){0.21}}
\multiput(30.7,89.49)(0.21,0.11){1}{\line(1,0){0.21}}
\multiput(30.91,89.6)(0.21,0.1){1}{\line(1,0){0.21}}
\multiput(31.12,89.7)(0.21,0.09){1}{\line(1,0){0.21}}
\multiput(31.33,89.78)(0.21,0.07){1}{\line(1,0){0.21}}
\multiput(31.54,89.86)(0.21,0.06){1}{\line(1,0){0.21}}
\multiput(31.75,89.91)(0.21,0.04){1}{\line(1,0){0.21}}
\multiput(31.96,89.96)(0.21,0.03){1}{\line(1,0){0.21}}
\multiput(32.18,89.98)(0.21,0.01){1}{\line(1,0){0.21}}
\put(32.39,90){\line(1,0){0.21}}
\multiput(32.61,90)(0.21,-0.01){1}{\line(1,0){0.21}}
\multiput(32.82,89.98)(0.21,-0.03){1}{\line(1,0){0.21}}
\multiput(33.04,89.96)(0.21,-0.04){1}{\line(1,0){0.21}}
\multiput(33.25,89.91)(0.21,-0.06){1}{\line(1,0){0.21}}
\multiput(33.46,89.86)(0.21,-0.07){1}{\line(1,0){0.21}}
\multiput(33.67,89.78)(0.21,-0.09){1}{\line(1,0){0.21}}
\multiput(33.88,89.7)(0.21,-0.1){1}{\line(1,0){0.21}}
\multiput(34.09,89.6)(0.21,-0.11){1}{\line(1,0){0.21}}
\multiput(34.3,89.49)(0.21,-0.13){1}{\line(1,0){0.21}}
\multiput(34.51,89.36)(0.21,-0.14){1}{\line(1,0){0.21}}
\multiput(34.72,89.22)(0.2,-0.15){1}{\line(1,0){0.2}}
\multiput(34.92,89.06)(0.2,-0.17){1}{\line(1,0){0.2}}
\multiput(35.12,88.9)(0.1,-0.09){2}{\line(1,0){0.1}}
\multiput(35.32,88.72)(0.1,-0.1){2}{\line(1,0){0.1}}
\multiput(35.52,88.52)(0.1,-0.1){2}{\line(0,-1){0.1}}
\multiput(35.71,88.31)(0.1,-0.11){2}{\line(0,-1){0.11}}
\multiput(35.9,88.09)(0.09,-0.12){2}{\line(0,-1){0.12}}
\multiput(36.09,87.86)(0.09,-0.12){2}{\line(0,-1){0.12}}
\multiput(36.28,87.61)(0.09,-0.13){2}{\line(0,-1){0.13}}
\multiput(36.46,87.36)(0.09,-0.14){2}{\line(0,-1){0.14}}
\multiput(36.64,87.09)(0.18,-0.28){1}{\line(0,-1){0.28}}
\multiput(36.82,86.8)(0.17,-0.29){1}{\line(0,-1){0.29}}
\multiput(36.99,86.51)(0.17,-0.31){1}{\line(0,-1){0.31}}
\multiput(37.16,86.2)(0.17,-0.32){1}{\line(0,-1){0.32}}
\multiput(37.33,85.89)(0.16,-0.33){1}{\line(0,-1){0.33}}
\multiput(37.49,85.56)(0.16,-0.34){1}{\line(0,-1){0.34}}
\multiput(37.65,85.22)(0.15,-0.35){1}{\line(0,-1){0.35}}
\multiput(37.8,84.87)(0.15,-0.36){1}{\line(0,-1){0.36}}
\multiput(37.95,84.52)(0.14,-0.37){1}{\line(0,-1){0.37}}
\multiput(38.1,84.15)(0.14,-0.38){1}{\line(0,-1){0.38}}
\multiput(38.24,83.77)(0.14,-0.39){1}{\line(0,-1){0.39}}
\multiput(38.37,83.38)(0.13,-0.4){1}{\line(0,-1){0.4}}
\multiput(38.5,82.99)(0.13,-0.4){1}{\line(0,-1){0.4}}
\multiput(38.63,82.58)(0.12,-0.41){1}{\line(0,-1){0.41}}
\multiput(38.75,82.17)(0.12,-0.42){1}{\line(0,-1){0.42}}
\multiput(38.87,81.75)(0.11,-0.43){1}{\line(0,-1){0.43}}
\multiput(38.98,81.32)(0.11,-0.44){1}{\line(0,-1){0.44}}
\multiput(39.08,80.89)(0.1,-0.44){1}{\line(0,-1){0.44}}
\multiput(39.18,80.44)(0.09,-0.45){1}{\line(0,-1){0.45}}
\multiput(39.28,80)(0.09,-0.45){1}{\line(0,-1){0.45}}
\multiput(39.37,79.54)(0.08,-0.46){1}{\line(0,-1){0.46}}
\multiput(39.45,79.08)(0.08,-0.47){1}{\line(0,-1){0.47}}
\multiput(39.53,78.62)(0.07,-0.47){1}{\line(0,-1){0.47}}
\multiput(39.6,78.14)(0.07,-0.48){1}{\line(0,-1){0.48}}
\multiput(39.67,77.67)(0.06,-0.48){1}{\line(0,-1){0.48}}
\multiput(39.73,77.19)(0.05,-0.48){1}{\line(0,-1){0.48}}
\multiput(39.78,76.71)(0.05,-0.49){1}{\line(0,-1){0.49}}
\multiput(39.83,76.22)(0.04,-0.49){1}{\line(0,-1){0.49}}
\multiput(39.87,75.73)(0.04,-0.49){1}{\line(0,-1){0.49}}
\multiput(39.91,75.24)(0.03,-0.49){1}{\line(0,-1){0.49}}
\multiput(39.94,74.74)(0.02,-0.5){1}{\line(0,-1){0.5}}
\multiput(39.96,74.25)(0.02,-0.5){1}{\line(0,-1){0.5}}
\multiput(39.98,73.75)(0.01,-0.5){1}{\line(0,-1){0.5}}
\multiput(39.99,73.25)(0.01,-0.5){1}{\line(0,-1){0.5}}

\linethickness{0.3mm}
\put(45,57.25){\line(0,1){0.5}}
\multiput(44.99,56.75)(0.01,0.5){1}{\line(0,1){0.5}}
\multiput(44.96,56.25)(0.02,0.5){1}{\line(0,1){0.5}}
\multiput(44.93,55.75)(0.04,0.5){1}{\line(0,1){0.5}}
\multiput(44.88,55.26)(0.05,0.5){1}{\line(0,1){0.5}}
\multiput(44.82,54.76)(0.06,0.49){1}{\line(0,1){0.49}}
\multiput(44.74,54.27)(0.07,0.49){1}{\line(0,1){0.49}}
\multiput(44.66,53.78)(0.09,0.49){1}{\line(0,1){0.49}}
\multiput(44.56,53.29)(0.1,0.49){1}{\line(0,1){0.49}}
\multiput(44.45,52.81)(0.11,0.48){1}{\line(0,1){0.48}}
\multiput(44.33,52.33)(0.12,0.48){1}{\line(0,1){0.48}}
\multiput(44.2,51.86)(0.13,0.48){1}{\line(0,1){0.48}}
\multiput(44.05,51.38)(0.14,0.47){1}{\line(0,1){0.47}}
\multiput(43.9,50.92)(0.16,0.47){1}{\line(0,1){0.47}}
\multiput(43.73,50.46)(0.17,0.46){1}{\line(0,1){0.46}}
\multiput(43.55,50)(0.18,0.45){1}{\line(0,1){0.45}}
\multiput(43.37,49.56)(0.09,0.22){2}{\line(0,1){0.22}}
\multiput(43.17,49.11)(0.1,0.22){2}{\line(0,1){0.22}}
\multiput(42.95,48.68)(0.11,0.22){2}{\line(0,1){0.22}}
\multiput(42.73,48.25)(0.11,0.21){2}{\line(0,1){0.21}}
\multiput(42.5,47.83)(0.12,0.21){2}{\line(0,1){0.21}}
\multiput(42.26,47.42)(0.12,0.21){2}{\line(0,1){0.21}}
\multiput(42.01,47.01)(0.13,0.2){2}{\line(0,1){0.2}}
\multiput(41.75,46.62)(0.13,0.2){2}{\line(0,1){0.2}}
\multiput(41.48,46.23)(0.14,0.19){2}{\line(0,1){0.19}}
\multiput(41.19,45.85)(0.14,0.19){2}{\line(0,1){0.19}}
\multiput(40.91,45.48)(0.14,0.18){2}{\line(0,1){0.18}}
\multiput(40.61,45.13)(0.15,0.18){2}{\line(0,1){0.18}}
\multiput(40.3,44.78)(0.1,0.12){3}{\line(0,1){0.12}}
\multiput(39.98,44.44)(0.11,0.11){3}{\line(0,1){0.11}}
\multiput(39.66,44.11)(0.11,0.11){3}{\line(0,1){0.11}}
\multiput(39.33,43.8)(0.11,0.11){3}{\line(1,0){0.11}}
\multiput(38.99,43.49)(0.11,0.1){3}{\line(1,0){0.11}}
\multiput(38.64,43.2)(0.17,0.15){2}{\line(1,0){0.17}}
\multiput(38.29,42.91)(0.18,0.14){2}{\line(1,0){0.18}}
\multiput(37.93,42.64)(0.18,0.14){2}{\line(1,0){0.18}}
\multiput(37.56,42.39)(0.18,0.13){2}{\line(1,0){0.18}}
\multiput(37.19,42.14)(0.19,0.12){2}{\line(1,0){0.19}}
\multiput(36.81,41.91)(0.19,0.12){2}{\line(1,0){0.19}}
\multiput(36.43,41.69)(0.19,0.11){2}{\line(1,0){0.19}}
\multiput(36.04,41.48)(0.19,0.1){2}{\line(1,0){0.19}}
\multiput(35.64,41.28)(0.2,0.1){2}{\line(1,0){0.2}}
\multiput(35.24,41.1)(0.2,0.09){2}{\line(1,0){0.2}}
\multiput(34.84,40.94)(0.4,0.17){1}{\line(1,0){0.4}}
\multiput(34.43,40.78)(0.41,0.15){1}{\line(1,0){0.41}}
\multiput(34.02,40.64)(0.41,0.14){1}{\line(1,0){0.41}}
\multiput(33.61,40.51)(0.41,0.13){1}{\line(1,0){0.41}}
\multiput(33.19,40.4)(0.42,0.11){1}{\line(1,0){0.42}}
\multiput(32.77,40.3)(0.42,0.1){1}{\line(1,0){0.42}}
\multiput(32.35,40.22)(0.42,0.09){1}{\line(1,0){0.42}}
\multiput(31.92,40.14)(0.42,0.07){1}{\line(1,0){0.42}}
\multiput(31.5,40.09)(0.43,0.06){1}{\line(1,0){0.43}}
\multiput(31.07,40.04)(0.43,0.04){1}{\line(1,0){0.43}}
\multiput(30.64,40.02)(0.43,0.03){1}{\line(1,0){0.43}}
\multiput(30.21,40)(0.43,0.01){1}{\line(1,0){0.43}}
\put(29.79,40){\line(1,0){0.43}}
\multiput(29.36,40.02)(0.43,-0.01){1}{\line(1,0){0.43}}
\multiput(28.93,40.04)(0.43,-0.03){1}{\line(1,0){0.43}}
\multiput(28.5,40.09)(0.43,-0.04){1}{\line(1,0){0.43}}
\multiput(28.08,40.14)(0.43,-0.06){1}{\line(1,0){0.43}}
\multiput(27.65,40.22)(0.42,-0.07){1}{\line(1,0){0.42}}
\multiput(27.23,40.3)(0.42,-0.09){1}{\line(1,0){0.42}}
\multiput(26.81,40.4)(0.42,-0.1){1}{\line(1,0){0.42}}
\multiput(26.39,40.51)(0.42,-0.11){1}{\line(1,0){0.42}}
\multiput(25.98,40.64)(0.41,-0.13){1}{\line(1,0){0.41}}
\multiput(25.57,40.78)(0.41,-0.14){1}{\line(1,0){0.41}}
\multiput(25.16,40.94)(0.41,-0.15){1}{\line(1,0){0.41}}
\multiput(24.76,41.1)(0.4,-0.17){1}{\line(1,0){0.4}}
\multiput(24.36,41.28)(0.2,-0.09){2}{\line(1,0){0.2}}
\multiput(23.96,41.48)(0.2,-0.1){2}{\line(1,0){0.2}}
\multiput(23.57,41.69)(0.19,-0.1){2}{\line(1,0){0.19}}
\multiput(23.19,41.91)(0.19,-0.11){2}{\line(1,0){0.19}}
\multiput(22.81,42.14)(0.19,-0.12){2}{\line(1,0){0.19}}
\multiput(22.44,42.39)(0.19,-0.12){2}{\line(1,0){0.19}}
\multiput(22.07,42.64)(0.18,-0.13){2}{\line(1,0){0.18}}
\multiput(21.71,42.91)(0.18,-0.14){2}{\line(1,0){0.18}}
\multiput(21.36,43.2)(0.18,-0.14){2}{\line(1,0){0.18}}
\multiput(21.01,43.49)(0.17,-0.15){2}{\line(1,0){0.17}}
\multiput(20.67,43.8)(0.11,-0.1){3}{\line(1,0){0.11}}
\multiput(20.34,44.11)(0.11,-0.11){3}{\line(1,0){0.11}}
\multiput(20.02,44.44)(0.11,-0.11){3}{\line(0,-1){0.11}}
\multiput(19.7,44.78)(0.11,-0.11){3}{\line(0,-1){0.11}}
\multiput(19.39,45.13)(0.1,-0.12){3}{\line(0,-1){0.12}}
\multiput(19.09,45.48)(0.15,-0.18){2}{\line(0,-1){0.18}}
\multiput(18.81,45.85)(0.14,-0.18){2}{\line(0,-1){0.18}}
\multiput(18.52,46.23)(0.14,-0.19){2}{\line(0,-1){0.19}}
\multiput(18.25,46.62)(0.14,-0.19){2}{\line(0,-1){0.19}}
\multiput(17.99,47.01)(0.13,-0.2){2}{\line(0,-1){0.2}}
\multiput(17.74,47.42)(0.13,-0.2){2}{\line(0,-1){0.2}}
\multiput(17.5,47.83)(0.12,-0.21){2}{\line(0,-1){0.21}}
\multiput(17.27,48.25)(0.12,-0.21){2}{\line(0,-1){0.21}}
\multiput(17.05,48.68)(0.11,-0.21){2}{\line(0,-1){0.21}}
\multiput(16.83,49.11)(0.11,-0.22){2}{\line(0,-1){0.22}}
\multiput(16.63,49.56)(0.1,-0.22){2}{\line(0,-1){0.22}}
\multiput(16.45,50)(0.09,-0.22){2}{\line(0,-1){0.22}}
\multiput(16.27,50.46)(0.18,-0.45){1}{\line(0,-1){0.45}}
\multiput(16.1,50.92)(0.17,-0.46){1}{\line(0,-1){0.46}}
\multiput(15.95,51.38)(0.16,-0.47){1}{\line(0,-1){0.47}}
\multiput(15.8,51.86)(0.14,-0.47){1}{\line(0,-1){0.47}}
\multiput(15.67,52.33)(0.13,-0.48){1}{\line(0,-1){0.48}}
\multiput(15.55,52.81)(0.12,-0.48){1}{\line(0,-1){0.48}}
\multiput(15.44,53.29)(0.11,-0.48){1}{\line(0,-1){0.48}}
\multiput(15.34,53.78)(0.1,-0.49){1}{\line(0,-1){0.49}}
\multiput(15.26,54.27)(0.09,-0.49){1}{\line(0,-1){0.49}}
\multiput(15.18,54.76)(0.07,-0.49){1}{\line(0,-1){0.49}}
\multiput(15.12,55.26)(0.06,-0.49){1}{\line(0,-1){0.49}}
\multiput(15.07,55.75)(0.05,-0.5){1}{\line(0,-1){0.5}}
\multiput(15.04,56.25)(0.04,-0.5){1}{\line(0,-1){0.5}}
\multiput(15.01,56.75)(0.02,-0.5){1}{\line(0,-1){0.5}}
\multiput(15,57.25)(0.01,-0.5){1}{\line(0,-1){0.5}}
\put(15,57.25){\line(0,1){0.5}}
\multiput(15,57.75)(0.01,0.5){1}{\line(0,1){0.5}}
\multiput(15.01,58.25)(0.02,0.5){1}{\line(0,1){0.5}}
\multiput(15.04,58.75)(0.04,0.5){1}{\line(0,1){0.5}}
\multiput(15.07,59.25)(0.05,0.5){1}{\line(0,1){0.5}}
\multiput(15.12,59.74)(0.06,0.49){1}{\line(0,1){0.49}}
\multiput(15.18,60.24)(0.07,0.49){1}{\line(0,1){0.49}}
\multiput(15.26,60.73)(0.09,0.49){1}{\line(0,1){0.49}}
\multiput(15.34,61.22)(0.1,0.49){1}{\line(0,1){0.49}}
\multiput(15.44,61.71)(0.11,0.48){1}{\line(0,1){0.48}}
\multiput(15.55,62.19)(0.12,0.48){1}{\line(0,1){0.48}}
\multiput(15.67,62.67)(0.13,0.48){1}{\line(0,1){0.48}}
\multiput(15.8,63.14)(0.14,0.47){1}{\line(0,1){0.47}}
\multiput(15.95,63.62)(0.16,0.47){1}{\line(0,1){0.47}}
\multiput(16.1,64.08)(0.17,0.46){1}{\line(0,1){0.46}}
\multiput(16.27,64.54)(0.18,0.45){1}{\line(0,1){0.45}}
\multiput(16.45,65)(0.09,0.22){2}{\line(0,1){0.22}}
\multiput(16.63,65.44)(0.1,0.22){2}{\line(0,1){0.22}}
\multiput(16.83,65.89)(0.11,0.22){2}{\line(0,1){0.22}}
\multiput(17.05,66.32)(0.11,0.21){2}{\line(0,1){0.21}}
\multiput(17.27,66.75)(0.12,0.21){2}{\line(0,1){0.21}}
\multiput(17.5,67.17)(0.12,0.21){2}{\line(0,1){0.21}}
\multiput(17.74,67.58)(0.13,0.2){2}{\line(0,1){0.2}}
\multiput(17.99,67.99)(0.13,0.2){2}{\line(0,1){0.2}}
\multiput(18.25,68.38)(0.14,0.19){2}{\line(0,1){0.19}}
\multiput(18.52,68.77)(0.14,0.19){2}{\line(0,1){0.19}}
\multiput(18.81,69.15)(0.14,0.18){2}{\line(0,1){0.18}}
\multiput(19.09,69.52)(0.15,0.18){2}{\line(0,1){0.18}}
\multiput(19.39,69.87)(0.1,0.12){3}{\line(0,1){0.12}}
\multiput(19.7,70.22)(0.11,0.11){3}{\line(0,1){0.11}}
\multiput(20.02,70.56)(0.11,0.11){3}{\line(0,1){0.11}}
\multiput(20.34,70.89)(0.11,0.11){3}{\line(1,0){0.11}}
\multiput(20.67,71.2)(0.11,0.1){3}{\line(1,0){0.11}}
\multiput(21.01,71.51)(0.17,0.15){2}{\line(1,0){0.17}}
\multiput(21.36,71.8)(0.18,0.14){2}{\line(1,0){0.18}}
\multiput(21.71,72.09)(0.18,0.14){2}{\line(1,0){0.18}}
\multiput(22.07,72.36)(0.18,0.13){2}{\line(1,0){0.18}}
\multiput(22.44,72.61)(0.19,0.12){2}{\line(1,0){0.19}}
\multiput(22.81,72.86)(0.19,0.12){2}{\line(1,0){0.19}}
\multiput(23.19,73.09)(0.19,0.11){2}{\line(1,0){0.19}}
\multiput(23.57,73.31)(0.19,0.1){2}{\line(1,0){0.19}}
\multiput(23.96,73.52)(0.2,0.1){2}{\line(1,0){0.2}}
\multiput(24.36,73.72)(0.2,0.09){2}{\line(1,0){0.2}}
\multiput(24.76,73.9)(0.4,0.17){1}{\line(1,0){0.4}}
\multiput(25.16,74.06)(0.41,0.15){1}{\line(1,0){0.41}}
\multiput(25.57,74.22)(0.41,0.14){1}{\line(1,0){0.41}}
\multiput(25.98,74.36)(0.41,0.13){1}{\line(1,0){0.41}}
\multiput(26.39,74.49)(0.42,0.11){1}{\line(1,0){0.42}}
\multiput(26.81,74.6)(0.42,0.1){1}{\line(1,0){0.42}}
\multiput(27.23,74.7)(0.42,0.09){1}{\line(1,0){0.42}}
\multiput(27.65,74.78)(0.42,0.07){1}{\line(1,0){0.42}}
\multiput(28.08,74.86)(0.43,0.06){1}{\line(1,0){0.43}}
\multiput(28.5,74.91)(0.43,0.04){1}{\line(1,0){0.43}}
\multiput(28.93,74.96)(0.43,0.03){1}{\line(1,0){0.43}}
\multiput(29.36,74.98)(0.43,0.01){1}{\line(1,0){0.43}}
\put(29.79,75){\line(1,0){0.43}}
\multiput(30.21,75)(0.43,-0.01){1}{\line(1,0){0.43}}
\multiput(30.64,74.98)(0.43,-0.03){1}{\line(1,0){0.43}}
\multiput(31.07,74.96)(0.43,-0.04){1}{\line(1,0){0.43}}
\multiput(31.5,74.91)(0.43,-0.06){1}{\line(1,0){0.43}}
\multiput(31.92,74.86)(0.42,-0.07){1}{\line(1,0){0.42}}
\multiput(32.35,74.78)(0.42,-0.09){1}{\line(1,0){0.42}}
\multiput(32.77,74.7)(0.42,-0.1){1}{\line(1,0){0.42}}
\multiput(33.19,74.6)(0.42,-0.11){1}{\line(1,0){0.42}}
\multiput(33.61,74.49)(0.41,-0.13){1}{\line(1,0){0.41}}
\multiput(34.02,74.36)(0.41,-0.14){1}{\line(1,0){0.41}}
\multiput(34.43,74.22)(0.41,-0.15){1}{\line(1,0){0.41}}
\multiput(34.84,74.06)(0.4,-0.17){1}{\line(1,0){0.4}}
\multiput(35.24,73.9)(0.2,-0.09){2}{\line(1,0){0.2}}
\multiput(35.64,73.72)(0.2,-0.1){2}{\line(1,0){0.2}}
\multiput(36.04,73.52)(0.19,-0.1){2}{\line(1,0){0.19}}
\multiput(36.43,73.31)(0.19,-0.11){2}{\line(1,0){0.19}}
\multiput(36.81,73.09)(0.19,-0.12){2}{\line(1,0){0.19}}
\multiput(37.19,72.86)(0.19,-0.12){2}{\line(1,0){0.19}}
\multiput(37.56,72.61)(0.18,-0.13){2}{\line(1,0){0.18}}
\multiput(37.93,72.36)(0.18,-0.14){2}{\line(1,0){0.18}}
\multiput(38.29,72.09)(0.18,-0.14){2}{\line(1,0){0.18}}
\multiput(38.64,71.8)(0.17,-0.15){2}{\line(1,0){0.17}}
\multiput(38.99,71.51)(0.11,-0.1){3}{\line(1,0){0.11}}
\multiput(39.33,71.2)(0.11,-0.11){3}{\line(1,0){0.11}}
\multiput(39.66,70.89)(0.11,-0.11){3}{\line(0,-1){0.11}}
\multiput(39.98,70.56)(0.11,-0.11){3}{\line(0,-1){0.11}}
\multiput(40.3,70.22)(0.1,-0.12){3}{\line(0,-1){0.12}}
\multiput(40.61,69.87)(0.15,-0.18){2}{\line(0,-1){0.18}}
\multiput(40.91,69.52)(0.14,-0.18){2}{\line(0,-1){0.18}}
\multiput(41.19,69.15)(0.14,-0.19){2}{\line(0,-1){0.19}}
\multiput(41.48,68.77)(0.14,-0.19){2}{\line(0,-1){0.19}}
\multiput(41.75,68.38)(0.13,-0.2){2}{\line(0,-1){0.2}}
\multiput(42.01,67.99)(0.13,-0.2){2}{\line(0,-1){0.2}}
\multiput(42.26,67.58)(0.12,-0.21){2}{\line(0,-1){0.21}}
\multiput(42.5,67.17)(0.12,-0.21){2}{\line(0,-1){0.21}}
\multiput(42.73,66.75)(0.11,-0.21){2}{\line(0,-1){0.21}}
\multiput(42.95,66.32)(0.11,-0.22){2}{\line(0,-1){0.22}}
\multiput(43.17,65.89)(0.1,-0.22){2}{\line(0,-1){0.22}}
\multiput(43.37,65.44)(0.09,-0.22){2}{\line(0,-1){0.22}}
\multiput(43.55,65)(0.18,-0.45){1}{\line(0,-1){0.45}}
\multiput(43.73,64.54)(0.17,-0.46){1}{\line(0,-1){0.46}}
\multiput(43.9,64.08)(0.16,-0.47){1}{\line(0,-1){0.47}}
\multiput(44.05,63.62)(0.14,-0.47){1}{\line(0,-1){0.47}}
\multiput(44.2,63.14)(0.13,-0.48){1}{\line(0,-1){0.48}}
\multiput(44.33,62.67)(0.12,-0.48){1}{\line(0,-1){0.48}}
\multiput(44.45,62.19)(0.11,-0.48){1}{\line(0,-1){0.48}}
\multiput(44.56,61.71)(0.1,-0.49){1}{\line(0,-1){0.49}}
\multiput(44.66,61.22)(0.09,-0.49){1}{\line(0,-1){0.49}}
\multiput(44.74,60.73)(0.07,-0.49){1}{\line(0,-1){0.49}}
\multiput(44.82,60.24)(0.06,-0.49){1}{\line(0,-1){0.49}}
\multiput(44.88,59.74)(0.05,-0.5){1}{\line(0,-1){0.5}}
\multiput(44.93,59.25)(0.04,-0.5){1}{\line(0,-1){0.5}}
\multiput(44.96,58.75)(0.02,-0.5){1}{\line(0,-1){0.5}}
\multiput(44.99,58.25)(0.01,-0.5){1}{\line(0,-1){0.5}}

\linethickness{0.3mm}
\put(50,39.8){\line(0,1){0.4}}
\multiput(49.98,39.4)(0.02,0.4){1}{\line(0,1){0.4}}
\multiput(49.94,38.99)(0.04,0.4){1}{\line(0,1){0.4}}
\multiput(49.88,38.59)(0.06,0.4){1}{\line(0,1){0.4}}
\multiput(49.8,38.2)(0.08,0.4){1}{\line(0,1){0.4}}
\multiput(49.69,37.8)(0.1,0.39){1}{\line(0,1){0.39}}
\multiput(49.57,37.41)(0.12,0.39){1}{\line(0,1){0.39}}
\multiput(49.43,37.02)(0.14,0.39){1}{\line(0,1){0.39}}
\multiput(49.27,36.64)(0.16,0.38){1}{\line(0,1){0.38}}
\multiput(49.1,36.27)(0.18,0.38){1}{\line(0,1){0.38}}
\multiput(48.9,35.9)(0.1,0.19){2}{\line(0,1){0.19}}
\multiput(48.68,35.53)(0.11,0.18){2}{\line(0,1){0.18}}
\multiput(48.45,35.18)(0.12,0.18){2}{\line(0,1){0.18}}
\multiput(48.2,34.83)(0.13,0.17){2}{\line(0,1){0.17}}
\multiput(47.93,34.49)(0.13,0.17){2}{\line(0,1){0.17}}
\multiput(47.64,34.15)(0.14,0.17){2}{\line(0,1){0.17}}
\multiput(47.34,33.83)(0.1,0.11){3}{\line(0,1){0.11}}
\multiput(47.02,33.52)(0.11,0.1){3}{\line(1,0){0.11}}
\multiput(46.69,33.22)(0.11,0.1){3}{\line(1,0){0.11}}
\multiput(46.34,32.93)(0.17,0.15){2}{\line(1,0){0.17}}
\multiput(45.98,32.65)(0.18,0.14){2}{\line(1,0){0.18}}
\multiput(45.6,32.38)(0.19,0.13){2}{\line(1,0){0.19}}
\multiput(45.21,32.13)(0.19,0.13){2}{\line(1,0){0.19}}
\multiput(44.81,31.89)(0.2,0.12){2}{\line(1,0){0.2}}
\multiput(44.39,31.66)(0.21,0.11){2}{\line(1,0){0.21}}
\multiput(43.97,31.44)(0.21,0.11){2}{\line(1,0){0.21}}
\multiput(43.53,31.24)(0.22,0.1){2}{\line(1,0){0.22}}
\multiput(43.08,31.05)(0.22,0.09){2}{\line(1,0){0.22}}
\multiput(42.63,30.88)(0.45,0.17){1}{\line(1,0){0.45}}
\multiput(42.17,30.72)(0.46,0.16){1}{\line(1,0){0.46}}
\multiput(41.7,30.58)(0.47,0.14){1}{\line(1,0){0.47}}
\multiput(41.22,30.45)(0.48,0.13){1}{\line(1,0){0.48}}
\multiput(40.74,30.34)(0.48,0.11){1}{\line(1,0){0.48}}
\multiput(40.25,30.24)(0.49,0.1){1}{\line(1,0){0.49}}
\multiput(39.75,30.16)(0.49,0.08){1}{\line(1,0){0.49}}
\multiput(39.26,30.1)(0.5,0.06){1}{\line(1,0){0.5}}
\multiput(38.76,30.05)(0.5,0.05){1}{\line(1,0){0.5}}
\multiput(38.25,30.02)(0.5,0.03){1}{\line(1,0){0.5}}
\multiput(37.75,30)(0.5,0.02){1}{\line(1,0){0.5}}
\put(37.25,30){\line(1,0){0.5}}
\multiput(36.75,30.02)(0.5,-0.02){1}{\line(1,0){0.5}}
\multiput(36.24,30.05)(0.5,-0.03){1}{\line(1,0){0.5}}
\multiput(35.74,30.1)(0.5,-0.05){1}{\line(1,0){0.5}}
\multiput(35.25,30.16)(0.5,-0.06){1}{\line(1,0){0.5}}
\multiput(34.75,30.24)(0.49,-0.08){1}{\line(1,0){0.49}}
\multiput(34.26,30.34)(0.49,-0.1){1}{\line(1,0){0.49}}
\multiput(33.78,30.45)(0.48,-0.11){1}{\line(1,0){0.48}}
\multiput(33.3,30.58)(0.48,-0.13){1}{\line(1,0){0.48}}
\multiput(32.83,30.72)(0.47,-0.14){1}{\line(1,0){0.47}}
\multiput(32.37,30.88)(0.46,-0.16){1}{\line(1,0){0.46}}
\multiput(31.92,31.05)(0.45,-0.17){1}{\line(1,0){0.45}}
\multiput(31.47,31.24)(0.22,-0.09){2}{\line(1,0){0.22}}
\multiput(31.03,31.44)(0.22,-0.1){2}{\line(1,0){0.22}}
\multiput(30.61,31.66)(0.21,-0.11){2}{\line(1,0){0.21}}
\multiput(30.19,31.89)(0.21,-0.11){2}{\line(1,0){0.21}}
\multiput(29.79,32.13)(0.2,-0.12){2}{\line(1,0){0.2}}
\multiput(29.4,32.38)(0.19,-0.13){2}{\line(1,0){0.19}}
\multiput(29.02,32.65)(0.19,-0.13){2}{\line(1,0){0.19}}
\multiput(28.66,32.93)(0.18,-0.14){2}{\line(1,0){0.18}}
\multiput(28.31,33.22)(0.17,-0.15){2}{\line(1,0){0.17}}
\multiput(27.98,33.52)(0.11,-0.1){3}{\line(1,0){0.11}}
\multiput(27.66,33.83)(0.11,-0.1){3}{\line(1,0){0.11}}
\multiput(27.36,34.15)(0.1,-0.11){3}{\line(0,-1){0.11}}
\multiput(27.07,34.49)(0.14,-0.17){2}{\line(0,-1){0.17}}
\multiput(26.8,34.83)(0.13,-0.17){2}{\line(0,-1){0.17}}
\multiput(26.55,35.18)(0.13,-0.17){2}{\line(0,-1){0.17}}
\multiput(26.32,35.53)(0.12,-0.18){2}{\line(0,-1){0.18}}
\multiput(26.1,35.9)(0.11,-0.18){2}{\line(0,-1){0.18}}
\multiput(25.9,36.27)(0.1,-0.19){2}{\line(0,-1){0.19}}
\multiput(25.73,36.64)(0.18,-0.38){1}{\line(0,-1){0.38}}
\multiput(25.57,37.02)(0.16,-0.38){1}{\line(0,-1){0.38}}
\multiput(25.43,37.41)(0.14,-0.39){1}{\line(0,-1){0.39}}
\multiput(25.31,37.8)(0.12,-0.39){1}{\line(0,-1){0.39}}
\multiput(25.2,38.2)(0.1,-0.39){1}{\line(0,-1){0.39}}
\multiput(25.12,38.59)(0.08,-0.4){1}{\line(0,-1){0.4}}
\multiput(25.06,38.99)(0.06,-0.4){1}{\line(0,-1){0.4}}
\multiput(25.02,39.4)(0.04,-0.4){1}{\line(0,-1){0.4}}
\multiput(25,39.8)(0.02,-0.4){1}{\line(0,-1){0.4}}
\put(25,39.8){\line(0,1){0.4}}
\multiput(25,40.2)(0.02,0.4){1}{\line(0,1){0.4}}
\multiput(25.02,40.6)(0.04,0.4){1}{\line(0,1){0.4}}
\multiput(25.06,41.01)(0.06,0.4){1}{\line(0,1){0.4}}
\multiput(25.12,41.41)(0.08,0.4){1}{\line(0,1){0.4}}
\multiput(25.2,41.8)(0.1,0.39){1}{\line(0,1){0.39}}
\multiput(25.31,42.2)(0.12,0.39){1}{\line(0,1){0.39}}
\multiput(25.43,42.59)(0.14,0.39){1}{\line(0,1){0.39}}
\multiput(25.57,42.98)(0.16,0.38){1}{\line(0,1){0.38}}
\multiput(25.73,43.36)(0.18,0.38){1}{\line(0,1){0.38}}
\multiput(25.9,43.73)(0.1,0.19){2}{\line(0,1){0.19}}
\multiput(26.1,44.1)(0.11,0.18){2}{\line(0,1){0.18}}
\multiput(26.32,44.47)(0.12,0.18){2}{\line(0,1){0.18}}
\multiput(26.55,44.82)(0.13,0.17){2}{\line(0,1){0.17}}
\multiput(26.8,45.17)(0.13,0.17){2}{\line(0,1){0.17}}
\multiput(27.07,45.51)(0.14,0.17){2}{\line(0,1){0.17}}
\multiput(27.36,45.85)(0.1,0.11){3}{\line(0,1){0.11}}
\multiput(27.66,46.17)(0.11,0.1){3}{\line(1,0){0.11}}
\multiput(27.98,46.48)(0.11,0.1){3}{\line(1,0){0.11}}
\multiput(28.31,46.78)(0.17,0.15){2}{\line(1,0){0.17}}
\multiput(28.66,47.07)(0.18,0.14){2}{\line(1,0){0.18}}
\multiput(29.02,47.35)(0.19,0.13){2}{\line(1,0){0.19}}
\multiput(29.4,47.62)(0.19,0.13){2}{\line(1,0){0.19}}
\multiput(29.79,47.87)(0.2,0.12){2}{\line(1,0){0.2}}
\multiput(30.19,48.11)(0.21,0.11){2}{\line(1,0){0.21}}
\multiput(30.61,48.34)(0.21,0.11){2}{\line(1,0){0.21}}
\multiput(31.03,48.56)(0.22,0.1){2}{\line(1,0){0.22}}
\multiput(31.47,48.76)(0.22,0.09){2}{\line(1,0){0.22}}
\multiput(31.92,48.95)(0.45,0.17){1}{\line(1,0){0.45}}
\multiput(32.37,49.12)(0.46,0.16){1}{\line(1,0){0.46}}
\multiput(32.83,49.28)(0.47,0.14){1}{\line(1,0){0.47}}
\multiput(33.3,49.42)(0.48,0.13){1}{\line(1,0){0.48}}
\multiput(33.78,49.55)(0.48,0.11){1}{\line(1,0){0.48}}
\multiput(34.26,49.66)(0.49,0.1){1}{\line(1,0){0.49}}
\multiput(34.75,49.76)(0.49,0.08){1}{\line(1,0){0.49}}
\multiput(35.25,49.84)(0.5,0.06){1}{\line(1,0){0.5}}
\multiput(35.74,49.9)(0.5,0.05){1}{\line(1,0){0.5}}
\multiput(36.24,49.95)(0.5,0.03){1}{\line(1,0){0.5}}
\multiput(36.75,49.98)(0.5,0.02){1}{\line(1,0){0.5}}
\put(37.25,50){\line(1,0){0.5}}
\multiput(37.75,50)(0.5,-0.02){1}{\line(1,0){0.5}}
\multiput(38.25,49.98)(0.5,-0.03){1}{\line(1,0){0.5}}
\multiput(38.76,49.95)(0.5,-0.05){1}{\line(1,0){0.5}}
\multiput(39.26,49.9)(0.5,-0.06){1}{\line(1,0){0.5}}
\multiput(39.75,49.84)(0.49,-0.08){1}{\line(1,0){0.49}}
\multiput(40.25,49.76)(0.49,-0.1){1}{\line(1,0){0.49}}
\multiput(40.74,49.66)(0.48,-0.11){1}{\line(1,0){0.48}}
\multiput(41.22,49.55)(0.48,-0.13){1}{\line(1,0){0.48}}
\multiput(41.7,49.42)(0.47,-0.14){1}{\line(1,0){0.47}}
\multiput(42.17,49.28)(0.46,-0.16){1}{\line(1,0){0.46}}
\multiput(42.63,49.12)(0.45,-0.17){1}{\line(1,0){0.45}}
\multiput(43.08,48.95)(0.22,-0.09){2}{\line(1,0){0.22}}
\multiput(43.53,48.76)(0.22,-0.1){2}{\line(1,0){0.22}}
\multiput(43.97,48.56)(0.21,-0.11){2}{\line(1,0){0.21}}
\multiput(44.39,48.34)(0.21,-0.11){2}{\line(1,0){0.21}}
\multiput(44.81,48.11)(0.2,-0.12){2}{\line(1,0){0.2}}
\multiput(45.21,47.87)(0.19,-0.13){2}{\line(1,0){0.19}}
\multiput(45.6,47.62)(0.19,-0.13){2}{\line(1,0){0.19}}
\multiput(45.98,47.35)(0.18,-0.14){2}{\line(1,0){0.18}}
\multiput(46.34,47.07)(0.17,-0.15){2}{\line(1,0){0.17}}
\multiput(46.69,46.78)(0.11,-0.1){3}{\line(1,0){0.11}}
\multiput(47.02,46.48)(0.11,-0.1){3}{\line(1,0){0.11}}
\multiput(47.34,46.17)(0.1,-0.11){3}{\line(0,-1){0.11}}
\multiput(47.64,45.85)(0.14,-0.17){2}{\line(0,-1){0.17}}
\multiput(47.93,45.51)(0.13,-0.17){2}{\line(0,-1){0.17}}
\multiput(48.2,45.17)(0.13,-0.17){2}{\line(0,-1){0.17}}
\multiput(48.45,44.82)(0.12,-0.18){2}{\line(0,-1){0.18}}
\multiput(48.68,44.47)(0.11,-0.18){2}{\line(0,-1){0.18}}
\multiput(48.9,44.1)(0.1,-0.19){2}{\line(0,-1){0.19}}
\multiput(49.1,43.73)(0.18,-0.38){1}{\line(0,-1){0.38}}
\multiput(49.27,43.36)(0.16,-0.38){1}{\line(0,-1){0.38}}
\multiput(49.43,42.98)(0.14,-0.39){1}{\line(0,-1){0.39}}
\multiput(49.57,42.59)(0.12,-0.39){1}{\line(0,-1){0.39}}
\multiput(49.69,42.2)(0.1,-0.39){1}{\line(0,-1){0.39}}
\multiput(49.8,41.8)(0.08,-0.4){1}{\line(0,-1){0.4}}
\multiput(49.88,41.41)(0.06,-0.4){1}{\line(0,-1){0.4}}
\multiput(49.94,41.01)(0.04,-0.4){1}{\line(0,-1){0.4}}
\multiput(49.98,40.6)(0.02,-0.4){1}{\line(0,-1){0.4}}

\linethickness{0.3mm}
\multiput(80,80)(0.12,-0.16){125}{\line(0,-1){0.16}}
\linethickness{0.3mm}
\multiput(95,60)(0.12,0.16){125}{\line(0,1){0.16}}
\linethickness{0.3mm}
\put(80,80){\line(1,0){30}}
\linethickness{0.3mm}
\put(95,40){\line(0,1){20}}
\linethickness{0.3mm}
\put(85,75){\line(1,0){20}}
\linethickness{0.3mm}
\put(85,80){\line(1,0){15}}
\linethickness{0.3mm}
\put(90,70){\line(1,0){10}}
\linethickness{0.3mm}
\put(85,77.5){\line(1,0){20}}
\linethickness{0.3mm}
\put(87.5,72.5){\line(1,0){15}}
\linethickness{0.3mm}
\put(92.5,67.5){\line(1,0){5}}
\linethickness{0.3mm}
\multiput(20,75)(1.31,-0.12){42}{\line(1,0){1.31}}
\put(75,70){\vector(1,-0){0.12}}
\linethickness{0.3mm}
\multiput(32.5,80)(1.01,-0.12){42}{\line(1,0){1.01}}
\put(75,75){\vector(1,-0){0.12}}
\linethickness{0.3mm}
\multiput(40,55)(0.71,0.12){42}{\line(1,0){0.71}}
\linethickness{0.3mm}
\multiput(70,60)(0.24,0.12){42}{\line(1,0){0.24}}
\put(80,65){\vector(2,1){0.12}}
\linethickness{0.3mm}
\multiput(70,60)(0.24,-0.12){63}{\line(1,0){0.24}}
\put(85,52.5){\vector(2,-1){0.12}}
\linethickness{0.3mm}
\multiput(40,40)(1.07,0.12){42}{\line(1,0){1.07}}
\put(85,45){\vector(1,0){0.12}}
\put(17,75){\makebox(0,0)[cc]{$U_\alpha$}}
\put(33,84){\makebox(0,0)[cc]{$U_\beta$}}
\put(23,55){\makebox(0,0)[cc]{$U_\gamma$}}
\put(40,35){\makebox(0,0)[cc]{$U_\delta$}}

\put(77.5,82.5){\makebox(0,0)[cc]{$v_\alpha$}}
\put(112,83){\makebox(0,0)[cc]{$v_\beta$}}
\put(99,59){\makebox(0,0)[cc]{$v_\gamma$}}
\put(95,36){\makebox(0,0)[cc]{$v_\delta$}}

\put(50,69){\makebox(0,0)[cc]{$\phi_\alpha$}}
\put(50,81){\makebox(0,0)[cc]{$\phi_\beta$}}
\put(55,53){\makebox(0,0)[cc]{$\phi_\gamma$}}
\put(60,38){\makebox(0,0)[cc]{$\phi_\delta$}}
\end{picture}

\setlength{\belowcaptionskip}{-25pt}
\medskip
\caption{A simplicial complex and a partition of unity for a cover by four sets}
\label{part45}
\end{center}
\end{figure}
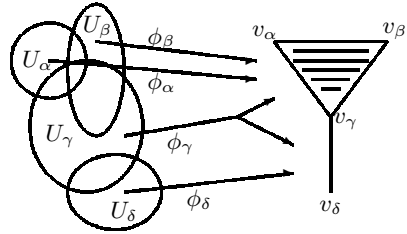

\begin{defin} \label{absc}
An {\em abstract simplicial complex} $(\mathcal{M},\mathcal{C})$ is a collection $\mathcal{C}$ of finite 
subsets  of a set $\mathcal{M}$ such that if $Y\subset X$ and $X \in \mathcal{C} $ then $Y\in \mathcal{C} $. 
\end{defin}

\begin{defin} \label{absc56}
A collection of subsets $U_\alpha\subset S$ (for $\alpha$ in some index set $\mathcal{M}$) is called a {\em cover}
if the union of all the $U_\alpha$ for $\alpha\in \mathcal{M}$ is equal to $S$. 
From such a cover we form an abstract simplicial complex, called the {\em nerve} 
of the cover, by 
\[
\mathcal{C} = \big\{ X\subset \mathcal{M} : \cap_{\alpha\in X} U_\alpha\neq\emptyset\big\}\ .
\]
\end{defin}

Thus $\{\alpha,\beta,\gamma\}\subset \mathcal{M} $ is in $\mathcal{C}$ precisely when $U_\alpha\cap U_\beta\cap U_\gamma$ is not empty. To see that $\mathcal{C}$ is an abstract simplicial complex we note that if 
$\{\alpha,\beta,\gamma\}$ is in $\mathcal{C}$ then we also must have $\{\alpha,\beta\}$ in $\mathcal{C}$ because if $U_\alpha\cap U_\beta\cap U_\gamma$ is not empty then also $U_\alpha\cap U_\beta$ is not empty. 

In Figure~\ref{part45} we have a space covered by four subsets $\{U_\alpha, U_\beta, U_\gamma, U_\delta\}$
and the corresponding abstract simplicial complex is
\[
\mathcal{C}=
\big\{ \{\alpha\},\{\beta\},\{\gamma\},\{\delta\},\{\alpha,\beta\},\{\alpha,\gamma\},\{\beta,\gamma\},\{\alpha,\beta,\gamma\},\{\gamma,\delta\}\big\}
\]

\begin{defin} \label{absmaps}
A map of abstract simplicial complexes $\Psi:(\mathcal{M},\mathcal{C})\to (\mathcal{M}',\mathcal{C}')$ is a function $\Psi:\mathcal{M}\to \mathcal{M}'$ so that on subsets
if $X\in \mathcal{C}$ then $\Psi X\in \mathcal{C}'$. 
\end{defin}

For example we can take $\mathcal{M}'=\{\delta,\zeta\}$ and
\[
\mathcal{C}'=
\big\{   \{\delta,\zeta\}, \{\delta\} ,\{\zeta\}
\big\}
\]
and a simplicial map $\Psi:\mathcal{M}\to \mathcal{M}'$ could be given by $\Psi(\alpha)=\Psi(\beta)=\Psi(\gamma)=\zeta$ and $\Psi(\delta)=\delta$.

We can use maps of abstract simplicial complexes to implement
hierarchies for information hiding or classification.

\begin{defin} \label{]refine}
A cover $\{W_k:k\in\mathcal{M}'\}$ of $S$ is a {\em refinement} of the cover $\{U_\alpha:\alpha\in\mathcal{M}\}$ if there is a map 
$\Psi:\mathcal{M}' \to \mathcal{M}$ so that $W_k\subset U_{\Psi(k)}$. It then follows that $\Psi$ is a map of abstract simplicial complexes from the nerve of the cover $\{W_k:k\in\mathcal{M}'\}$ to the nerve of $\{U_\alpha:\alpha\in\mathcal{M}\}$. 
\end{defin}

\subsection{Realisation of abstract simplicial complexes in $\mathbb{R}^{n}$}\label{real_simplicial_complexes}

The familiar $xy$ plane for 2-dimensional geometry is called $\mathbb{R}^2$ as it uses two copies of the real numbers $\mathbb{R}$. Its elements are ordered pairs $(x,y)$ for $x$ and $y$ real numbers. We have a basis $e_1=(1,0)$ and $e_2=(0,1)$ and can write any point in the plane as $(x,y)=x\,e_1+y\,e_2$. Similarly for 3-dimensional space $\mathbb{R}^3$ we have points $(x,y,z)$, and if we use new basis elements $e_1=(1,0,0)$, $e_2=(0,1,0)$ and $e_3=(0,0,1)$ we can write $(x,y,z)=x\,e_1+y\,e_2+z\,e_3$. 

In (for example) $\mathbb{R}^3$ we have points which we call $0$-simplices, e.g., $e_1$ and $e_2$. The 1-simplices are lines connecting the 0-simplices, e.g., $x\,e_1+y\,e_2$ for real $x,y\ge 0$ with $x+y=1$. The 2-simplices are triangles spanned by three vertices, e.g., $x\,e_1+y\,e_2+z\,e_3$ for real $x,y,z\ge 0$ with $x+y+z=1$. We extend this to be 3-simplices being tetrahedra, etc. A 3-simplex is visualised in Figure~\ref{part45} where the vertices are labelled $\{\alpha,\beta,\gamma\}$.

An $n$-simplex will have \textit{faces} which are simplices bounded by subsets of its vertices. Thus a 3-simplex will have one 3-face (itself), four 2-faces which are 2-simplices, six 1-faces which are 1-simplices and four 0-faces which are 0-simplices (vertices). In 
 Figure~\ref{part45}  we the 2-simplex  $\{\alpha,\beta,\gamma\}$ has a 1-face labelled by $\{\alpha,\gamma\}$ and a 0-face labelled by $\{\alpha\}$. 
 
We can use this to define a higher dimensional analogue of a planar graph called a simplicial complex -- a graph is an example of a simplicial complex which only contains 0-simplices and 1-simplices: 

\begin{defin} \label{realsc}
A {\em simplicial complex} is a collection of simplices in some space so that the face of any simplex in the collection is also in the collection, and the intersection of any two simplices is a face of both of them.
\end{defin}

Figure~\ref{part45} gives an example of a simplicial complex which is a union of a 2-simplex and a 1-simplex. These simplices intersect at the common 1-face $e_\gamma$. 

There is a construction which will give a simplicial complex for every abstract simplicial complex in a functorial manner. In general, this construction uses very high dimensional spaces, which is usually not necessary in practice, but it simplifies the theory. Just as many (but not all) graphs can be drawn in 2-dimensional space we can often draw a simplicial complex in a dimension much smaller than that used in Proposition~\ref{Real_representation}. The large space
 $\mathbb{R}^\mathcal{M}$ is a vector space with basis $e_\alpha$ for all $\alpha\in \mathcal{M}$.

\begin{propos}\label{Real_representation}
To every abstract simplicial complex $(\mathcal{M},\mathcal{C})$ (as in Definition~\ref{absc}) is associated its standard realisation $\Delta_\mathcal{C} \subset \mathbb{R}^\mathcal{M}$, as a simplicial complex. 
The simplex spanned by $X\in \mathcal{C}$ is
\[
\Delta_X=\Big\{ \sum_{\alpha\in X} \lambda_\alpha  \, e_\alpha : 
\lambda_\alpha\in[0,1],\  \sum_{\alpha\in X} \lambda_\alpha=1\Big\}. 
\] 
Further, a map of abstract simplicial complexes $\Psi:(\mathcal{M},\mathcal{C})\to (\mathcal{M}',\mathcal{C}')$ can be extended to a map of their realisations
as $\Delta_\Psi:\Delta_\mathcal{C} \to \Delta_\mathcal{C'} $ by defining
\[
\Delta_\Psi\Big(  \sum_{\alpha\in X} \lambda_\alpha  \, e_\alpha  \Big) = \sum_{\alpha\in X} \lambda_\alpha  \, e_{\Psi(\alpha)}.
\]
\end{propos}
\noindent\textbf{Proof:}\quad 
The simplex $\Delta_X$ is a $(|X|-1)$-simplex where $|X|$ is the size of $X$, and if $Y\subset X$ then $\Delta_Y$ is a face of $\Delta_X$. 
Then $\Delta_X\cap \Delta_Z=\Delta_{X\cap Z}$ and Definition~\ref{realsc} is seen to be satisfied. \qquad$\square$

\subsection{Covers and partitions of unity}\label{partitions}

Having created the geometric realisation of the simplex we are now able to use the categorisation of the data in the state space $S$ according to a cover $\{ U_\alpha : \alpha\in\mathcal{M} \}$ and its associated abstract simplicial complex $\mathcal{C}$, to create a geometric visualisation in $\Delta_\mathcal{C}$. To do this we need an appropriate map from $S$ to $\mathcal{C}$ that contains the information $\phi_\alpha(s)$ which tells us how much we need to be concerned about $\alpha$ when in state $s\in S$ on a scale from 0 to 1.

\begin{defin} \label{realpu}
A {\em partition of unity} for the cover $\{U_\alpha\subset S\,|\, \alpha\in\mathcal{M}\}$ is a function $\phi_\alpha:S\to[0,1]$ for every $\alpha\in\mathcal{M}$ such that 

(1)\quad
if $\phi_\alpha(s)\neq 0$ then $s\in U_\alpha$;

(2)\quad $\sum_{\alpha\in\mathcal{M}}\phi_\alpha(s)=1$ for all $s\in S$. 

\noindent We then have a function $\phi:S\to \Delta_{\mathcal{C}}$ given by
$$\phi(s)= \sum_{\alpha\in\mathcal{M}} \phi_\alpha(s)\,e_\alpha.$$

\end{defin}
Figure~\ref{part45} visualises a simplicial complex and partition of unity for a cover by four sets.
Note that the triangle in  Figure~\ref{part45} is shaded to form a 2-simplex precisely because $U_\alpha\cap U_\beta\cap U_\gamma$ is not empty.  In specific circumstances we can impose extra conditions on $\phi$, e.g., continuity or computability.

\subsection{Belief} \label{bel99}

 It is important to distinguish between belief in something and a probability that that thing occurs \cite{Shaf,DemSchBel}.
 On being told by a completely reliable source that a person has a car which has a single colour, we could form two statements, 
\begin{center}
$B=$`the car is blue' and $NB=$`the car is not blue'.
\end{center}

The combination $B$ or $NB$ we would consider to have a belief of 1, but in the absence of any other information we would have no basis to assign a non-zero belief value to the separate statement $B$ or to the statement $NB$. We could randomly assign belief values to each which added up to 1, but this would only undermine the idea that \textit{belief ought be assigned according to evidence}. Sensibly our belief $\mathrm{Bel}(\{B,NB\})$ in ($B$ or $NB$) is strictly greater than the sum of $\mathrm{Bel}(\{B\})$ and $\mathrm{Bel}(\{NB\})$. Such a conclusion means that we are not dealing with a probability distribution.

\begin{definition}\label{beldef}
  We consider beliefs in a finite set $X$ of statements.
A {\em generalised belief function} on $X$ is a function 
$$\mathrm{Bel}:P(X)\to [0,1],$$ 
where $P(X)$ is the set of subsets of $X$. Furthermore, for the empty set $\mathrm{Bel}(\emptyset)=0$ and
the sets satisfy the `super-additivity' property
\begin{eqnarray} \label{superadd}
\mathrm{Bel}(Y\cup Z)+\mathrm{Bel}(Y\cap Z)\ge \mathrm{Bel}(Y)+\mathrm{Bel}(Z)\quad\mathrm{for\ all}\ Y,Z\subset X.
\end{eqnarray}
 We denote by $\mathcal{B}(X)$ the set of belief functions on $X$. 
\end{definition}
 
We shall say that a belief function is {\em normalised} if $\mathrm{Bel}(X)=1$. It is important to note that we do \textit{not} impose this normalisation condition -- hence our word \textit{generalised} above. 
Thus, we reserve the right to believe that the set of presented alternatives $X$ may be incomplete.

The `super-additivity' property is simply justified in that, in addition to those people who believe in $Y$ and those who believe in $Z$, there may be people who just believe that at least one is true.

\subsection{Simplicial complexes and graphs} \label{gra7}
 
Graphs are a standard data structure in computer science. We claim that for certain purposes simplicial complexes are preferable. These purposes are related to the reasons why simplicial complexes were invented by topologists: In describing covers of spaces by subsets, these subsets can model multiple objectives or circumstances which can occur simultaneously. This intrinsically higher dimensional picture does not easily reduce to a graph. To see this, first consider this straightforward construction.

In transforming a simplicial complex to a graph, 

(i) each mode/face $f$ of the complex becomes a vertex $v$; and 

(ii) each pair of nested mode/faces  $f_1 \subset f_2$ of the complex becomes an edge $v_1 \to v_2$.

In summary:
\begin{lemma}\label{complex_to_graph}
An $n$-simplex contains $2^{n+1}-1$ faces. To make a representation in a graph, the resulting graph would have $2^{n+1}-1$ vertices.
\end{lemma}

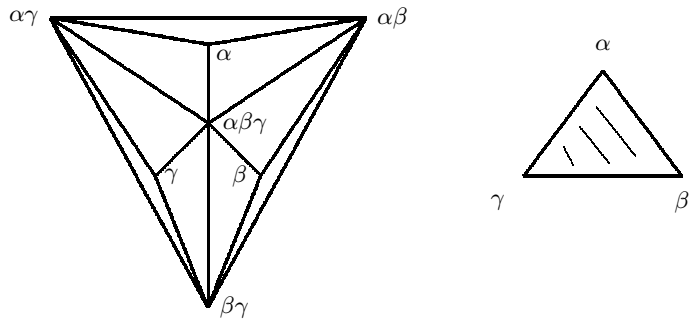
\begin{figure}[htbp]
\begin{center}
 \unitlength 0.9 mm
 
 \unitlength 0.7 mm
\begin{picture}(140,60)(0,0)
\linethickness{0.3mm}
\put(50,40){\line(0,1){15}}
\linethickness{0.3mm}
\multiput(50,40)(0.12,-0.12){83}{\line(1,0){0.12}}
\linethickness{0.3mm}
\multiput(40,30)(0.12,0.12){83}{\line(1,0){0.12}}
\put(75,50){\makebox(0,0)[cc]{}}

\linethickness{0.3mm}
\multiput(60,30)(0.12,0.18){167}{\line(0,1){0.18}}
\linethickness{0.3mm}
\multiput(50,55)(0.71,0.12){42}{\line(1,0){0.71}}
\linethickness{0.3mm}
\multiput(20,60)(0.71,-0.12){42}{\line(1,0){0.71}}
\linethickness{0.3mm}
\multiput(20,60)(0.12,-0.18){167}{\line(0,-1){0.18}}
\linethickness{0.3mm}
\multiput(40,30)(0.12,-0.3){83}{\line(0,-1){0.3}}
\linethickness{0.3mm}
\multiput(50,5)(0.12,0.3){83}{\line(0,1){0.3}}
\put(57,40){\makebox(0,0)[cc]{$\alpha\beta\gamma$}}

\put(85,60){\makebox(0,0)[cc]{$\alpha\beta$}}

\put(15,60){\makebox(0,0)[cc]{$\alpha\gamma$}}

\put(55,5){\makebox(0,0)[cc]{$\beta\gamma$}}

\put(53,53){\makebox(0,0)[cc]{$\alpha$}}

\put(56,30){\makebox(0,0)[cc]{$\beta$}}

\put(43,30){\makebox(0,0)[cc]{$\gamma$}}

\linethickness{0.3mm}
\multiput(50,40)(0.18,0.12){167}{\line(1,0){0.18}}
\linethickness{0.3mm}
\put(50,5){\line(0,1){35}}
\linethickness{0.3mm}
\multiput(20,60)(0.18,-0.12){167}{\line(1,0){0.18}}
\linethickness{0.3mm}
\multiput(110,30)(0.12,0.16){125}{\line(0,1){0.16}}
\linethickness{0.3mm}
\multiput(125,50)(0.12,-0.16){125}{\line(0,-1){0.16}}
\linethickness{0.3mm}
\put(110,30){\line(1,0){30}}
\put(125,55){\makebox(0,0)[cc]{$\alpha$}}

\put(140,25){\makebox(0,0)[cc]{$\beta$}}

\put(105,25){\makebox(0,0)[cc]{$\gamma$}}

\linethickness{0.05mm}
\multiput(123.75,43.12)(0.12,-0.15){63}{\line(0,-1){0.15}}
\linethickness{0.05mm}
\multiput(120.62,39.38)(0.12,-0.15){47}{\line(0,-1){0.15}}
\linethickness{0.05mm}
\multiput(117.5,35.62)(0.12,-0.23){16}{\line(0,-1){0.23}}
\linethickness{0.3mm}
\put(20,60){\line(1,0){60}}
\linethickness{0.3mm}
\multiput(50,5)(0.12,0.22){250}{\line(0,1){0.22}}
\linethickness{0.3mm}
\multiput(20,60)(0.12,-0.22){250}{\line(0,-1){0.22}}
\end{picture}

\setlength{\belowcaptionskip}{-15pt}
\medskip
\caption{Comparing a single 2-simplex (right) and its graph (left)}
\label{grcomp}
\end{center}
\end{figure}

In Figure~\ref{grcomp} we compare a simple  2-simplex $\alpha\beta\gamma$ to its graph, where we take the sub-simplices of $\alpha\beta\gamma$ and connect the intersections (= possible mode transitions) by a line in the graph. This shows the considerable increase in complexity on constructing the graph, even for one of the simplest simplicial complexes. The graphical representation of the simplicial complex in Figure~\ref{sectri} would be much more complicated. 

Not only is the graph more complicated, but it is much more difficult to interpret. A point on the 2-simplex (the filled in triangle) carries a large amount of information about the state of the system in how close to an edge or vertex it is. All points in the interior of the triangle correspond to the single vertex 
$\alpha\beta\gamma$ in the graph, so a position in a vertex of the graph loses this information. The discontinuous motion from vertex to vertex in the graph lacks any explainability or predictive value. This is in contrast to the continuous motion of a point around the 2-simplex, where at least a guess may be made about the likely future behaviour of the system.

\end{document}